\documentclass[aps, twocolumn]{revtex4}
\usepackage{graphicx}

\newcommand{\znco}{\mbox{$\rm ZnCr_2O_4$}}
\newcommand{\zcgo}{\mbox{$\rm ZnCr_{2-2x}Ga_{2x}O_4$}}
\newcommand{\gapone}{\mbox{$\rm ZnCr_{1.9}Ga_{0.1}O_4$}}
\newcommand{\gapsix}{\mbox{$\rm ZnCr_{1.4}Ga_{0.6}O_4$}}

\newcommand{\neel}{\mbox{Ne\'{e}l}}
\begin{document}

\title{\bf  N\'{e}el to Spin-Glass-like Phase
Transition versus Dilution
in Geometrically Frustrated $\rm \bf ZnCr_{2-2x}Ga_{2x}O_4$.}
\author{S.-H. Lee$^1$,
W. Ratcliff II$^2$,
Q. Huang$^{2}$, T. H.
Kim$^{3}$, and S-W. Cheong$^{4}$ }
\address{$^{1}$ Department of Physics, University of Virginia, Charlottesville, VA 22904}
\address{$^{2}$
NIST Center for Neutron Research, National Institute of Standards
and Technology, Gaithersburg, Maryland
20899}
\address{$^3$Department of Physics and Nanosciences, EWHA Womans University, Seoul 120-750, Korea}
\address{$^4$Department of Physics and Astronomy, Rutgers University,
Piscataway, New Jersey 08854}

\begin{abstract}
\znco~ undergoes a first order spin-Peierls-like phase transition
at 12.5 K from a cubic spin liquid phase to a tetragonal \neel~
state.\cite{zcoprl} Using powder diffraction and single crystal 
polarized neutron scattering, we determined the complex spin structure
of the \neel~ phase. This phase consisted of several magnetic domains
with different characteristic wave vectors. This indicates that the
tetragonal phase of \zcgo~ is very close to a critical point surrounded
by many different \neel~ states. We have also studied, using elastic
and inelastic neutron scattering techniques, the effect of nonmagnetic
dilution on magnetic correlations in \zcgo~ (x=0.05 and 0.3). For
x=0.05, the magnetic correlations do not change qualitatively from
those in the pure material, except that the phase transition becomes
second order. For x= 0.3, the spin-spin correlations become short range.
Interestingly, the spatial correlations of the frozen spins in the
x=0.3 material are the same as those of the fluctuating moments in
the pure and the weakly diluted materials. 
\end{abstract}

\pacs{PACS numbers: 76.50.+g, 
 75.40.Gb, 
  75.50.Ee}

\maketitle

\section{Introduction}

There has been a long standing fascination in the physics community
with placing antiferromagnetically coupled spins on lattices with 
triangular motifs.\cite{anderson,villain} In two dimensions, one
can consider the simple triangular lattice. When classical XY spins
are placed on this lattice, it orders at zero temperature. For many
years, the excitations above this ground state provided a playground 
for studying topological excitations. These chiral excitations destroy
the ground state at finite temperatures and experimental investigations
of these systems are still underway. One can increase the complexity 
of this problem by considering the case of the even less constrained
system of corner sharing triangles, the kagome lattice.\cite{ramirez,obradors,aprscgo90,cb90,scgoeuro}
There have been many theoretical questions about the nature of the 
ground state \cite{park,senthil,ran,ryu} even for classical spins
on this lattice.\cite{chubukov,huse,mila98} Experimentally, good
experimental realizations of this system have been hard to obtain.
Recently, however, single crystals of Fe-jarosite that realize the 
kagome lattice with classical (S = 5/2) spins were grown and detailed
neutron scattering studies have been performed.\cite{matan} 
More recently, ZnCu$_{3}$(OH)$_{6}$Cl$_{2}$
was found as a good model system for the quantum kagome antiferromagnet.\cite{shores,helton,ofer,mendels,shlnmat}
Unfortuately, it turned out that there is 5-10\% site switching of
Zn$^{2+}$ and Cu$^{2+}$ ions,\cite{shlnmat} which obscures the
quantum physics of the kagome antiferromagnet.

In three dimensions, when spins form a network of corner sharing tetrahedra, 
it leads to a macroscopically degenerate ground state for classical
as well as quantum spins.\cite{cana98,moes98} Theoretically novel
low temperature properties are expected to appear. For example, quantum
spin liquid phases, fractional excitations, or non-Ginzburg-Landau 
phase transitions. Experimentally, spinels AB$_{2}$O$_{4}$ have
attracted lots of attention because the B sublattice forms a network
of corner-sharing tetrahedra. In the spinel, the B site cations are
octahedrally coordinated by six oxygens and neighboring BO$_{6}$ 
octahedra share an edge. Thus, when the B site is occupied by a transition
metal ion with $t_{2g}$ electrons, the system can realize the simple
and most frustrating Heisenburg spin Hamiltonian, $H=J\sum{\bf S}_{i}\cdot{\bf S}_{j}$ 
with uniform nearest neighbor interactions.

ACr$_{2}$O$_{4}$ (A = Zn\cite{zcoprl}, Cd\cite{cdcr2o4}, Hg\cite{hgcr2o4})
realizes the most frustrating Hamiltonian because the $t_{2g}$ orbital
of the Cr$^{3+}$ $(3d^{3})$ is half filled and the nearest neighbor 
interactions due to the direct overlap of the neighboring $t_{2g}$
orbitals are dominant and spatially uniform.\cite{good} In comparison,
in the case of AV$_{2}$O$_{4}$ where the V$^{2+}(3d^{2})$ ion has
an orbital degeneracy, a Jahn-Teller distortion can occur at low temperatures, 
which makes the vanadates effectively one-dimensional spin chain systems.\cite{tsunetsugu,znv2o4,oleg_znv2o4,khomskii}
Several novel discoveries have been made in ACr$_{2}$O$_{4}$. For
instance, collective excitations of local \textit{antiferromagnetic} 
hexagonal spins were found in the spin liquid phase of \znco~that
embody the zero-energy excitations amongst the degenerate ground states.\cite{zconature,foroleg}
Unfortunately, the lattice of ACr$_{2}$O$_{4}$ is not infinitely 
firm and it distorts at low temperatures to lift the magnetic frustration.
The novel three-dimensional spin-Peierls phase transition, i.e., the
lattice instability driven by magnetic interactions, occurs and drives 
the system into a \neel~state. The lattice distortion can occur
in different forms, depending on details of the crystal environment:
tetragonal $I\bar{4}m2$ symmetry for ZnCr$_{2}$O$_{4}$\cite{acr2o4_cryst},
tetragonal $I4_{1}/amd$ for CdCr$_{2}$O$_{4}$\cite{acr2o4_cryst},
and orthorhombic $Fddd$ for HgCr$_{2}$O$_{4}$\cite{hueda,hgcr2o4}. When an external
magnetic field is applied to the \neel~state, the half-magnetization
plateau states appear in CdCr$_{2}$O$_{4}$\cite{hueda_prl} and 
HgCr$_{2}$O$_{4}$\cite{hueda,hgcr2o4} due to the field-induced
lattice instability\cite{penc,balents}.

\begin{figure}[!htb]
\begin{center}
\includegraphics[width=0.9\linewidth]{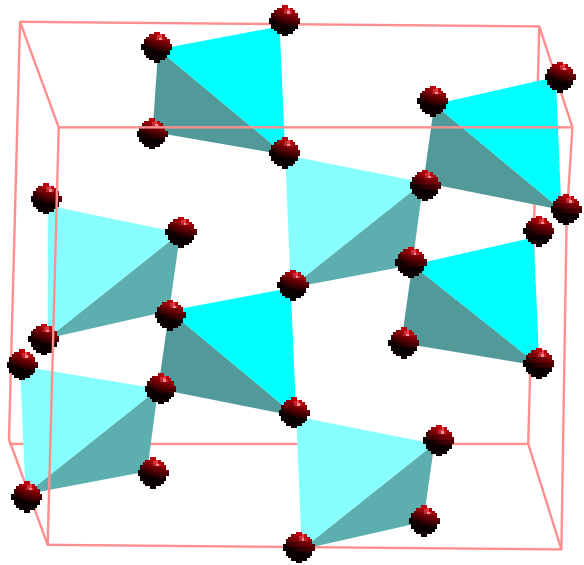}
\caption{
Octahedral B-sites of spinel $\rm AB_2O_4$ form a network of corner-sharing
tetrahedra.}
\end{center}
\end{figure}


In this paper, we investigated the nature of the 3D spin-Peierls transition in the chromite by performing elastic and inelastic neutron scattering
measurements on \zcgo~ for x=0, 0.05, and 0.3.  Our principal results are the following. For x=0, the
N\'{e}el state has four characteristic wave vectors, ${\bf
k}=(1,0,0), (\frac{1}{2},\frac{1}{2},\frac{1}{2}),
 (1,0,\frac{1}{2})$ and $(\frac{1}{2},\frac{1}{2},0)$.\cite{ks} The large
size of the magnetic unit cell (64 Cr$^{3+}$ ions) has made it
difficult to uniquely determine the spin structure of this system.
We have determined the spin structure, employing powder
diffraction, single crystal polarized neutron diffraction data and
a systematic group theoretical approach\cite{izyumov}.  We find
that the system is composed of three types of domains whose
relative fractions vary from sample to sample.  The dominant domain is a multi-k structure with
${\bf k}=(1,0,\frac{1}{2})$ and $(\frac{1}{2},\frac{1}{2},0)$. This spin
structure is coplanar and noncollinear with spins pointing along
either the $a$ or $b$ axis with each tetraheron having two pairs
of antiparallel spins to have zero net moment. The ${\bf
k}=(\frac{1}{2},\frac{1}{2},\frac{1}{2})$ domain has a rather
simple spin structure.  The spinel lattice can be decomposed into
alternating kagome and triangular layers when viewed along the
$<111>$ direction.  In this spin structure, the spins in the
kagome layer order in the "q=0" configuration.  The spins in the
triangular layer point along the $<111>$ direction and are
parallel within a layer.  Spins in alternating layers are
antiparallel. The ${\bf k}=(1,0,0)$
domain has a collinear spin structure with spins parallel to the
z-axis, as in ZnV$_2$O$_4$.\cite{znv2o4}  In each tetrahedron the net spin is zero. 

The effect of site disorder on the
magnetic correlations and phase transition in \znco~ by doping
nonmagnetic Ga ions into Cr sites has been previously studied by Fiorani {\it et al.} using bulk property measurements and neutron powder diffractions in the
\zcgo~ series\cite{fiorani84}. Fig. 2 shows the phase diagram that
they have constructed from the measurements. N\'{e}el phase
survives up to $x\sim 0.2$. For $0.2<x<x_c$ with $1-x_c=0.390(3)$
being the percolation threshold for the corner-sharing
tetrahedra\cite{percol}, the system exhibits spin-glass-like
properties in bulk susceptibility measurements. However, the
nonlinear susceptibility of \zcgo~(x=0.2) does not display the
divergence expected of an ordinary spin glass\cite{hammann}. This
suggests that the low temperature phase is not an ordinary spin
glass.

\begin{figure}[!htb]
\begin{center}
\includegraphics[width=2.6in,angle=90]{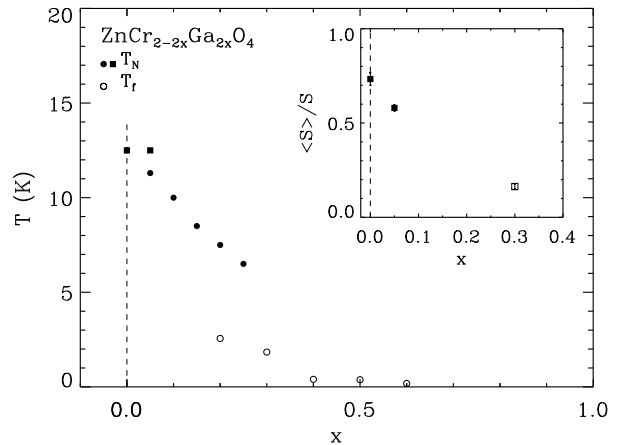}
\caption{
Phase diagram for \zcgo~. N\'{e}el temperature $T_N$ and spin
freezing temperature $T_f$ represented by circles are the data
obtained by bulk susceptibility measurements reported in Ref.
\cite{fiorani84}. $T_N$ represented by squares and $\frac{<S>}{S}$
in the inset are obtained by our neutron scattering measurements,
which is discussed in the Sections IV and V.}
\end{center}
\end{figure}

Our data show that in the weakly diluted \zcgo~(x=0.05) N\'{e}el ordering occurs with
the same spin structure as the parent compound.  However, the
ordering now develops gradually and the phase transition becomes
second order.  The appearance of the \neel~ ordering and the
cubic-to-tetragonal structural phase now also proceed in a second
order manner.  This consanguinity of the order of the structural
phase transition and the appearance of long-range magnetic
ordering supports our interpretation that the phase transition is
magnetically driven.

For \zcgo~(x=0.3), the magnetic long range order is replaced by
static short range order even though the Cr concentration,
$1-x=0.7$, is well above the percolation threshold, 0.390.
Interestingly, the spatial correlations of the frozen spins in the
spin-glass sample are the same as those of the fluctuating moments
present in the pure and weakly diluted materials. Magnetic neutron
scattering intensity, $\tilde{I} (Q)$, vanishes as $Q\to 0$ and
has a broad peak at $Q_c \simeq 1.5$ \AA$^{-1}$ with
full-width-of-half-maximum (FWHM) of $\kappa=0.48(5)$ \AA$^{-1}$.
This indicates that fundamental spin degree of freedom in the
corner-sharing tetrahedra involves an {\it antiferromagnetic} hexagonal spin loop with zero
net moment, which distinguish the geometrically frustrated magnet
from an ordinary spin glass.

The structure of this paper is the following:  In section II, we
describe the experimental details of material synthesis and the
neutron scattering techniques that were used.  In Section  III, we
explain the determination of the spin structure of
$ZnCr_{2}O_{4}$. In Section IV, we discuss inelastic neutron
scattering data on the material and how the spin freezing and
short range spin correlations in the diluted compound resemble
those in other frustrated magnets.  This paper concludes with a
discussion and summary in section V.

\section{Experimental Details}

Three 20 g powder samples of \znco~, one \gapone~ sample and
\gapsix~  sample were prepared by the standard solid state
reaction method with stoichiometric amounts of Cr$_2$O$_3$,
Ga$_2$O$_3$ and ZnO in air. Neutron powder diffraction
measurements performed on the samples at the National Institute of
Standards and Technology (NIST) BT1 diffractometer  show that the
samples were stoichiometric single phase spinels with the
exception of one \znco~ sample (sample 2) which had a minority
phase of 1\% f.u. unreacted $\rm Cr_2O_3$. The results of the
structural refinement are summarized in Table I.  The \znco~
samples will be denoted by sample 1, sample 2, and sample 3 in
this paper.

A 0.1 g single crystal of \znco~ was grown by the chemical
transport method and used for polarized neutron diffraction
measurements at the NIST cold neutron triple-axis spectrometer,
SPINS. Spectrometer configuration was
guide-PG(002)-Be-Pol.-40$^{'}$-Samp.-Flip.-Pol.-40$^{'}$-PG(002)-Det.
The sample was mounted such that the scattering plane were the
(hk0) and (h0l) zones due to twinning.  A vertical guide field was
applied. The polarization efficiency was determined by measuring
the scattering intensities of a nuclear (2,2,0) Bragg peak with
the flipper on and off. Correction for the finite polarizing
efficiency, 0.85(1), was made\cite{polcor}.

For inelastic neutron scattering
measurements on powder samples, we utilized
a multiplexing detection system of SPINS consisting of a flat analyzer and
a position-sensitive detector.
The details of the experimental setup are reported elsewhere\cite{zcoprl}.
  High angle backgrounds were
measured by defocusing the analyzer while low angle backgrounds from
air scattering were measured by extracting the sample from the
cryostat.  The absolute efficiency of the detection system was measured
using incoherent elastic scattering from vanadium and
nuclear Bragg peaks from the samples.
The corresponding correction factor was applied to the background
subtracted data to obtain normalized measurements of the magnetic
scattering cross section\cite{kcodsoprb}.

\section{Antiferromagnetic long-range order in \znco}

\subsection{Powder diffraction data}

%
\begin{figure}[!htb]
\begin{center}
\includegraphics[width=3.4in,angle=0]{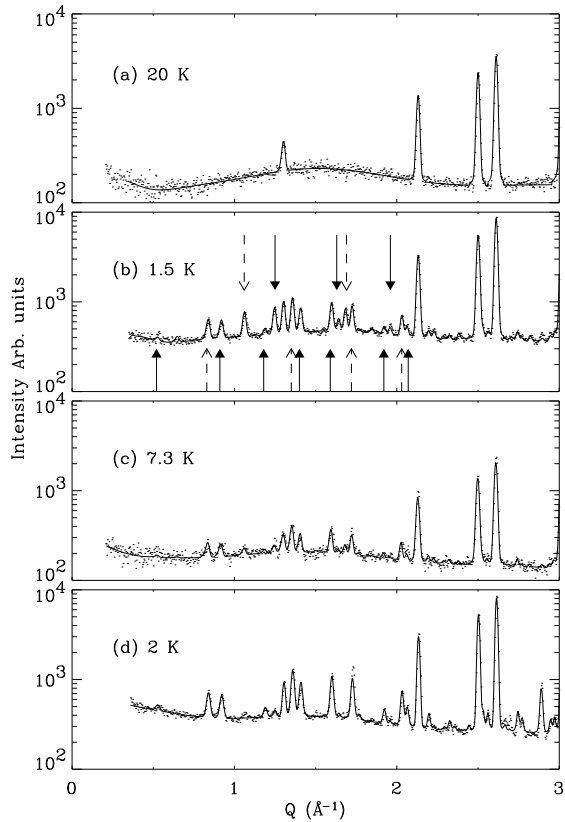}
\caption{
Powder
diffraction data from $\rm ZnCr_2O_4$ at 25 K and 2 K taken on the
powder diffractometer BT 1 at NIST. The line through the data in
(a) shows the Rietveld fit to the crystal structure of which
parameters are shown in Table 1. The line in (b) is the fit to the
crystal structure and the magnetic structure that are explained in
the text for sample 1. The upward pointing arrows with plain line
are magnetic reflections which belong to ${\bf
k}=(\frac{1}{2},\frac{1}{2},0)$. From the left, they are
$(\frac{1}{2},\frac{1}{2},0), (\frac{1}{2},\frac{1}{2},1),
(\frac{1}{2},\bar{\frac{3}{2}},0),
(\bar{\frac{1}{2}},\frac{3}{2},\bar{1}),
(\frac{1}{2},\frac{1}{2},2)$ and $(\frac{3}{2},\frac{3}{2},0),
(\bar{\frac{1}{2}},\frac{3}{2},2)$ and
$(\frac{1}{2},\frac{5}{2},0)$, and $(\frac{1}{2},\frac{5}{2},1)$.
The upward arrows with dashed line are magnetic reflections which
belong to ${\bf k}=(1,0,\frac{1}{2})$. From the left, they are
$(1,0,\frac{1}{2}), (1,0,\frac{3}{2}), (1,2,\frac{1}{2})$, and
$(2,1,\frac{3}{2})$ and $(1,0,\frac{5}{2})$. The downward pointing
plain arrows are from the ${\bf
k}=(\frac{1}{2},\frac{1}{2},\frac{1}{2})$ family of magnetic
reflections. From the left, they are
$(\bar{\frac{1}{2}},\bar{\frac{1}{2}},\bar{\frac{3}{2}})$,
$(\frac{3}{2},\frac{3}{2},\frac{1}{2})$, and
$(\frac{1}{2},\frac{5}{2},\frac{1}{2})$.  The downward pointing
dashed arrows are from the ${\bf k}=(1,0,0)$ family of magnetic
reflections.  From the left, they are $(1,1,0)$, $(2,1,0)$, and
$(2,1,1)$.  Figures (c) and (d) show the data and fits for samples
(2) and (3) respectively, also discussed in the text.}
\end{center}
\end{figure}

Fig. 3 (a) shows $T = 25 K > T_N$ diffraction data from \znco~ with the Rietveld
fit superimposed. In addition to the nuclear Bragg reflections,
there is a broad peak centered at $Q\sim 1.5$ \AA$^{-1}$. This broad
peak is due to dynamic spin fluctuations and will be discussed
in the Section IV.
Below $T_c$, the broad peak weakens and
magnetic Bragg reflections appear, indicating
a long-range magnetic ordering.
These diffraction patterns are consistent with those observed
previously from $\rm ZnFe_{0.1}Cr_{1.9}O_4$\cite{oles70}.
Indexing these magnetic reflections indicates that the magnetic
unit cell consists of four chemical formula units (64 magnetic
Cr$^{3+}$ ions) which
can be characterized by four wave vectors,
${\bf k}=(\frac{1}{2},\frac{1}{2},0)$, $(1,0,\frac{1}{2})$, $(\frac{1}{2},\frac{1}{2},\frac{1}{2})$,
$(1,0,0)$
It is impossible to uniquely determine spin structure
for a system with such a large magnetic unit cell
only from its powder diffraction pattern.
Ol\'{e}s proposed an inplane
spin structure for $\rm ZnFe_{0.1}Cr_{1.9}O_4$\cite{oles70}
and Shaked {\it et al.} a non-inplane structure
for $\rm MgCr_2O_4$\cite{shaked70}.
Apparently, as we will show in Section III. C,
there are numerous spin structures that can explain
the neutron powder diffraction  data equally well.
To obtain more restrictive information for the spin structure,
we have performed polarized neutron diffraction on a single crystal of \znco.  Our polarization
study focuses on the ${\bf k}=(\frac{1}{2},\frac{1}{2},0)$ and $(1,0,\frac{1}{2})$ family of
magnetic reflections.

\subsection{Polarized neutron diffraction data from a single crystal}

This material undergoes a cubic-to-tetragonal structural phase transition
with $c<a$ at $T_N=12.5$ K\cite{zcoprl}.
Because of tetragonal twinning, below $T_N$
a wave vector transfer ${\bf Q}=(Q_x,Q_y,Q_z)$ in the laboratory coordinate
system
represents $(Q_x/a^*,Q_y/a^*,Q_z/c^*)\equiv (h,k,l)$ and
$(Q_x/a^*,Q_y/c^*,Q_z/a^*)\equiv (h,l,k)$
in different crystal twin domains.
In the configuration with a vertical guide field,
the non-spin-flip (NSF) and spin-flip (SF) scattering
cross sections, $\sigma_{NSF}$ and $\sigma_{SF}$ become\cite{moon,lovesey}
\begin{eqnarray}
\sigma_{NSF} & = & \sigma_N+\sigma_M^z \nonumber \\
\sigma_{SF} & = &
\sigma_{M\perp}^x+\sigma_{M\perp}^y.
\end{eqnarray}
Here $\sigma_{N}$ is structural scattering cross section,
and $\sigma_{M}$ is the magnetic scattering cross section,
$\sigma_M \propto  (1-\hat{{\bf Q}}\cdot\widehat{S})
|F_M({\bf Q})|^2$. In $\sigma_{NSF}$, we neglected the interference term
between the nuclear and the magnetic scattering amplitude
because the reflections considered here are either purely nuclear
or purely magnetic.
$F_M({\bf Q})$ is the magnetic structure factor
$F_M({\bf Q})=\frac{1}{2}gF({\bf Q})\Sigma_d <S_d>e^{i{\bf Q}\cdot \vec{d}}$
where $F(\vec{\tau})$ is the magnetic form
factor of Cr$^{3+}$\cite{xraytables}.
The perpendicular sign in $\sigma_{M\perp}$ is to emphasize
that only the spin components perpendicular to the wave vector
transfer ${\bf Q}$ contribute to scattering.
Fig. 3 (a) shows the NSF and SF scattering intensities
obtained at a nuclear (2,2,0) Bragg reflection. Only NSF scattering
is expected for this nuclear Bragg reflection and the contribution
in the SF channel due to the contamination from
the incomplete instrumental polarization efficiency
of 0.85(1).
For magnetic Bragg reflections that belong to
the ${\bf k}=(\frac{1}{2},\frac{1}{2},0)$ family,
Eq. (1) becomes
\begin{eqnarray}
\sigma_{NSF} & = & \sigma_M^c \nonumber \\
\sigma_{SF} & = &
\sigma_{M\perp}^a+\sigma_{M\perp}^b.
\end{eqnarray}

\begin{figure}[!htb]
\begin{center}
\includegraphics[width=2.6in,angle=90]{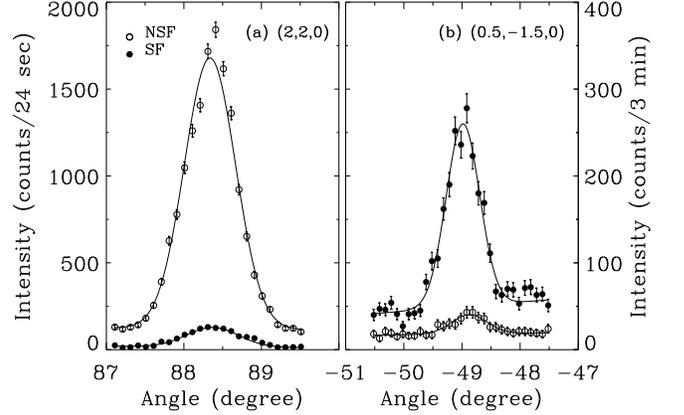}
\caption{
Rocking scan through (a) a nuclear Bragg reflection at (2,2,0)
and (b) a magnetic Bragg reflection at $(\frac{1}{2},-\frac{3}{2},0)$
obtained from a single crystal of \znco~ at $T=1.7$ K.
Open circles are the NSF data and filled ones are the SF data.}
\end{center}
\end{figure}

As shown in Fig. 4 (b), The magnetic $(\frac{1}{2},\bar{\frac{3}{2}},0)$
Bragg reflection has dominantly SF intensity and
a weak signal in the
NSF channel. The weak NSF intensity  is due to contamination from
incomplete polarization.
We investigated five magnetic Bragg reflections:
$(\frac{1}{2},\frac{1}{2},0), (\frac{1}{2},\bar{\frac{3}{2}},0),
(\frac{3}{2},\frac{3}{2},0), (\frac{1}{2},\frac{5}{2},0)$,
and $(\bar{\frac{3}{2}},\frac{5}{2},0)$.
After the correction for the incomplete polarization,
at these reflections only SF scattering is present.
This means that the spins are in the ab-plane, $\sigma_M^c=0$.

\begin{figure}[!htb]
\begin{center}
\includegraphics[width=2.6in,angle=90]{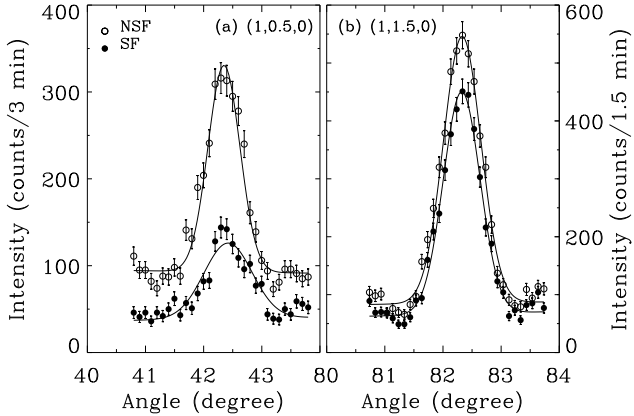}
\caption{
Rocking scan through magnetic Bragg reflections (a) at $(1,\frac{1}{2},0)$
and (b) at $(1,\frac{3}{2},0)$
obtained from a single crystal of \znco~ at $T=1.7$ K.
Open circles are the NSF data and filled ones are the SF data.}
\end{center}
\end{figure}

For the magnetic Bragg reflections that belong to ${\bf k}=(1,0,\frac{1}{2})$,
there are NSF as well as SF scattering. The ratio of SF to NSF scattering
intensity, $\sigma_{SF}/\sigma_{NSF}$,
is different at different $Q$, as shown in Fig. 5.
Table I lists the ratios for three different reflections.
The ratio, $\sigma_{SF}/\sigma_{NSF}$, increases as $l$ increases.
This information places a strict restriction
on any possible spin configuration for \znco.
For these reflections, Eq. (1) becomes
\begin{eqnarray}
\sigma_{NSF} & = & \sigma_M^b \nonumber \\
\sigma_{SF} & = &
\sigma_{M\perp}^a+\sigma_{M\perp}^c.
\end{eqnarray}

Therefore, using $\sigma_M^c=0$,
\begin{equation}
\frac{\sigma_{SF}}{\sigma_{NSF}} = \left( 1-\frac{h^2}{h^2+l^2}\right)
 \cdot \frac{ |F_M^a|^2}{ |F_M^b|^2}.
\end{equation}
The fact that the measured $\sigma_{SF}/\sigma_{NSF}$
follows $1-\frac{h^2}{h^2+l^2}$ within the experimental accuracy
indicates $|F_M^a|=|F_M^b|$ for the ${\bf k}=(1,0,\frac{1}{2})$ reflections.

\subsection{Group theoretical approach to determination of spin structure}

We have employed a group theoretical approach developed by
Izyumov {\it et al.}\cite{izyumov} to determine the spin structure.
The basic idea of the method
is that any magnetic structure with a characteristic wave vector ${\bf k}$
can be expanded in terms of basis functions, $\psi^{{\bf k}_L}_{\lambda}$,
of irreducible representations
of the spin space of the crystal $G_{\bf k}$ which is a
subgroup of the crystal space group $G$\cite{izyumov},
\begin{eqnarray}
S^{\{{\bf k}\}}_{0j} & = & \sum_L S_{0j}^{{\bf k}_L} \nonumber \\
 & = & \sum_L \sum_\lambda C^{{\bf k}_L}_{\lambda} \psi^{{\bf k}_L}_{\lambda}.
\end{eqnarray}
Here $0j$ represents a magnetic ion at site $j$ in zeroth primitive cell.
$L$ runs over the arms of the star ${\bf k}_L$ and $\lambda$ over irreducible
representations of the star arm ${\bf k}_L$.
The star of a wave vector ${\bf k}$, $\{ {\bf k} \}$, is the set
of nonequivalent vectors that can be obtained by acting on ${\bf k}$
with an element of the crystal space group $g\in G$.
For instance, the star $\{{\bf k}\}=\{(1,0,\frac{1}{2})\}$ has
six arms: ${\bf k}_L=(1,0,\frac{1}{2}),
(\bar{1},0,\bar{\frac{1}{2}}), (1,\frac{1}{2},0),
(\bar{1},\bar{\frac{1}{2}},0), (\frac{1}{2},0,1)$, and
$(\bar{\frac{1}{2}},0,\bar{1})$.
Once $S^{\{{\bf k}\}}_{0j}$ is determined,
all spins at other primitive cells, $S_{nj}$, can be derived
by\cite{izyumov}
\begin{equation}
S_{nj} = \sum_L {\rm exp} (i {\bf k}\cdot {\bf t}_n) S_{0j}^{{\bf k}_L}
\end{equation}
where ${\bf t}_n$ is the translation vector for the $n$ cell
from the zeroth primitive cell.
How to obtain the basis functions of the irreducible representations,
$\psi^{{\bf k}_L}_{\lambda}$  for a given  $G_{\bf k}$ has been
explained in a great detail in a book by Izyumov {\it et al.}\cite{izyumov}
and will not be repeated here.
The basis functions $\psi^{{\bf k}_L}_{\lambda}$ for
${\bf k}=(\frac{1}{2},\frac{1}{2},0)$ and ${\bf k}=(1,0,\frac{1}{2})$, are
listed in Table II and III respectively.
Note that $\psi^{{\bf k}_L}_{\lambda}$ are complex
but their simple superposition at the two arms
$C_1 \psi^{{\bf k_1} \tau}+C_2 \psi^{{\bf k_2} \tau}$
with ${\bf k}_2=-{\bf k}_1$ can generate
a real function, provided that the coefficients $C_1$ and $C_2$ are
appropriately selected.
Table IV and V list such superpositions for
${\bf k}=(\frac{1}{2},\frac{1}{2},0)$ and
${\bf k}=(1,0,\frac{1}{2})$, respectively.
The superpositions of two irreducible represtations
,$C_1 \psi^{{\bf k_1} \tau}+C_2 \psi^{{\bf k_2} \tau}$,
do not yield nonzero spins for all 16 magnetic ions
in a chemical unit cell. Instead
the superposition yields four nonzero spins for
${\bf k}=(\frac{1}{2},\frac{1}{2},0)$
and eight nonzero spins for ${\bf k}=(1,0,\frac{1}{2})$ in a chemical unit cell.
This means that to put all 16 nonzero spins into a chemical unit cell
we have to consider at least four of
the $C_1 \psi^{{\bf k_1} \tau}+C_2 \psi^{{\bf k_2} \tau}$ listed in Table III
for ${\bf k}=(\frac{1}{2},\frac{1}{2},0)$
and at least two of $C_1 \psi^{{\bf k_1} \tau}+C_2 \psi^{{\bf k_2} \tau}$
listed in Table IV for ${\bf k}=(1,0,\frac{1}{2})$.
Obviously the number of such combinations is very large.
To narrow down the possible spin structure,
we used the constraints
that were obtained from polarized neutron diffraction data:
(1) spins are in the ab-plane, $S_c=0$, and
(2) $|F_M^a|=|F_M^b|$ for the $(1,0,\frac{1}{2})$ family
reflections.  We also assumed that
(3) all spins have the same magnitude and that (4) all tetrahedra have
zero net spin.

\begin{figure}[!htb]
\begin{center}
\includegraphics[width=3.4in]{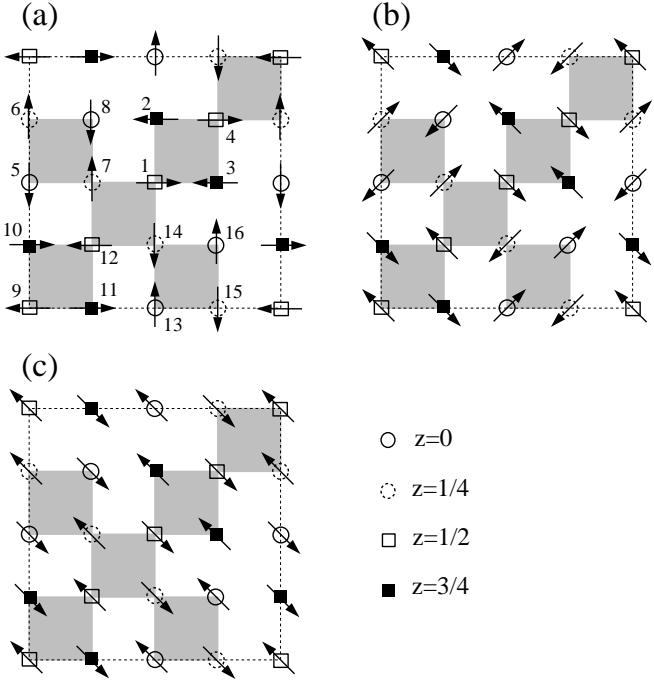}
\caption{
Prototypes of possible spin structures in a chemical unit cell
for ${\bf k}=(1,0,\frac{1}{2})$
which satisfy the conditions described in the text:
(a) $\psi^{{\bf k_1} \tau_{11}}+ \psi^{{\bf k_2} \tau_{12}}
+ i \psi^{{\bf k_1} \tau_{11}^{'}} - i  \psi^{{\bf k_2} \tau_{12}^{'}}$,
(b) $(1+i) \psi^{{\bf k_1} \tau_{11}}+ (1-i) \psi^{{\bf k_2} \tau_{12}}
+ (-1+i) \psi^{{\bf k_1} \tau_{11}^{'}} + (-1-i)  \psi^{{\bf k_2} \tau_{12}^{'}}$,
and (c) $\psi^{{\bf k_1} \tau_{11}}+ \psi^{{\bf k_2} \tau_{12}}
+ i \psi^{{\bf k_1} \tau_{11}^{'}} - i  \psi^{{\bf k_2} \tau_{12}^{'}}
- i \psi^{{\bf k_1} \tau_{21}} - i  \psi^{{\bf k_2} \tau_{22}}
-  \psi^{{\bf k_1} \tau_{21}^{'}} +  \psi^{{\bf k_2} \tau_{22}^{'}}$.
Here ${\bf k_1}=(1,0,\frac{1}{2})$
and ${\bf k_2}=(\bar{1},0,\bar{\frac{1}{2}})$.
Magnetic unit cell is doubled along the c-axis and spins change the sign
in the chemical unit cell displaced by (0,0,1). A shaded square represents
a tetrahedron formed by four Cr$^{3+}$ ions.
Symbols represent z-coordinates of the magnetic Cr$^{3+}$ ions.}
\end{center}
\end{figure}

Since for this domain, there are two characteristic wave vectors
involved in the N\'{e}el state of \znco, we rewrite Eq. (5) to
separate $S^{\{{\bf k}\}}_{0j}$ into two components;
\begin{eqnarray}
S^{\{{\bf k}\}}_{0j} & =
& \sum_L S_{0j}^{{\bf k}_L(1,0,\frac{1}{2})}
+ \sum_L S_{0j}^{{\bf k}_L(\frac{1}{2},\frac{1}{2},0)} \\ \nonumber
& \equiv & S^{\{{\bf k}(1,0,\frac{1}{2})\}}_{0j}
+ S^{\{{\bf k}(\frac{1}{2},\frac{1}{2},0)\}}_{0j}.
\end{eqnarray}
Note that $S^{\{{\bf k}(1,0,\frac{1}{2})\}}$
contribute only to the ${\bf k}=(1,0,\frac{1}{2})$
family reflections
and $S^{\{{\bf k}(\frac{1}{2},\frac{1}{2},0)\}}$
only to the ${\bf k}=(\frac{1}{2},\frac{1}{2},0)$ reflections.
Therefore we can obtain the two components separately.

\begin{figure}[!htb]
\begin{center}
\includegraphics[width=3.4in]{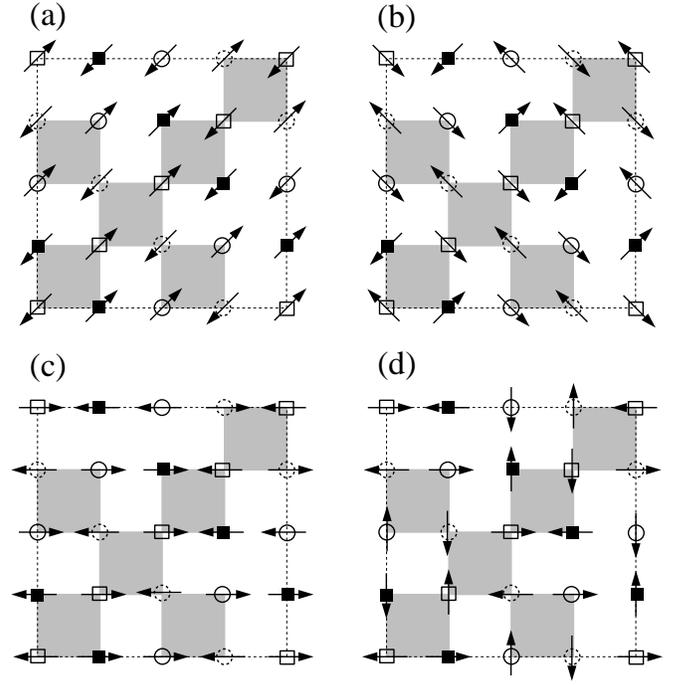}
\caption{
Prototypes of possible spin structures in a chemical unit cell for
${\bf k}=(\frac{1}{2},\frac{1}{2},0)$
in which satisfy the constraints described in the text:
(a) $\psi^{{\bf k_1} \tau_{1}^{'}}+ i \psi^{{\bf k_2} \tau_{1}^{'}}
+ (1+i) \psi^{{\bf k_1} \tau_{2}} + (1-i)  \psi^{{\bf k_2} \tau_{2}}
-i \psi^{{\bf k_1} \tau_{2}^{'}} -  \psi^{{\bf k_2} \tau_{2}^{'}}$,
(b) $\psi^{{\bf k_1} \tau_{1}^{'}}+ i \psi^{{\bf k_2} \tau_{1}^{'}}
+ (1+i) \psi^{{\bf k_1} \tau_{4}} + (1-i)  \psi^{{\bf k_2} \tau_{4}}
-i \psi^{{\bf k_1} \tau_{4}^{'}} -  \psi^{{\bf k_2} \tau_{4}^{'}}$,
(c) $i \psi^{{\bf k_1} \tau_{1}^{'}}+ \psi^{{\bf k_2} \tau_{1}^{'}}
+ (1-i) \psi^{{\bf k_1} \tau_{2}} + (1+i)  \psi^{{\bf k_2} \tau_{2}}
- \psi^{{\bf k_1} \tau_{2}^{'}} - i \psi^{{\bf k_2} \tau_{2}^{'}}
+ \psi^{{\bf k_1} \tau_{3}^{'}}+ i \psi^{{\bf k_2} \tau_{3}^{'}}
+ (1+i) \psi^{{\bf k_1} \tau_{4}} + (1-i)  \psi^{{\bf k_2} \tau_{4}}
- i \psi^{{\bf k_1} \tau_{4}^{'}} -  \psi^{{\bf k_2} \tau_{4}^{'}}$,
and (d) $i \psi^{{\bf k_1} \tau_{1}^{'}}+ \psi^{{\bf k_2} \tau_{1}^{'}}
+ (1-i) \psi^{{\bf k_1} \tau_{2}} + (1+i)  \psi^{{\bf k_2} \tau_{2}}
- \psi^{{\bf k_1} \tau_{2}^{'}} - i \psi^{{\bf k_2} \tau_{2}^{'}}
-i \psi^{{\bf k_1} \tau_{3}^{'}}- \psi^{{\bf k_2} \tau_{3}^{'}}
+ (1-i) \psi^{{\bf k_1} \tau_{4}} + (1+i)  \psi^{{\bf k_2} \tau_{4}}
- \psi^{{\bf k_1} \tau_{4}^{'}} - i \psi^{{\bf k_2} \tau_{4}^{'}}$.
Here ${\bf k_1}=(\frac{1}{2},\frac{1}{2},0)$
and ${\bf k_2}=(\bar{\frac{1}{2}},\bar{\frac{1}{2}},0)$.
The magnetic unit cell is doubled along the a- and the b-axes,
and spins change sign
in the chemical unit cell displaced by (1,0,0) or (0,1,0).
Symbols representing  z-coordinates of the magnetic Cr$^{3+}$ ions
are the same as those in Fig. 6.}
\end{center}
\end{figure}

First, let us consider the ${\bf k}=(1,0,\frac{1}{2})$ reflections.
We examined all possible combinations of the superpositions
listed in Table V and found 24 different spin configurations
which can be divided into three categories shown in Fig. 5.
Fig. 6 (a) shows a non-collinear spin configuration in which spins are
along either a-axis or b-axis, Fig. 6 (b) shows a non-collinear
spin configuration with spins along $(1,\bar{1},0)$ or $(\bar{1},1,0)$,
and Fig. 6 (c) shows a collinear spin configuration along $(\bar{1},1,0)$.
A collinear spin configuration along a-axis or b-axis is
ruled out by the constraint $|F_M^a|=|F_M^b|$.

Fig. 8 shows four prototypes of ${\bf k}=(\frac{1}{2},\frac{1}{2},0)$
spin configurations in which all tetrahedra satisfy the antiferromagnetic
constraint to have zero net moment:
(100) type collinear and noncollinear spin configurations
and (110) type collinear and noncollinear spin configurations.

\begin{figure}[!htb]
\begin{center}
\includegraphics[width=3.4in]{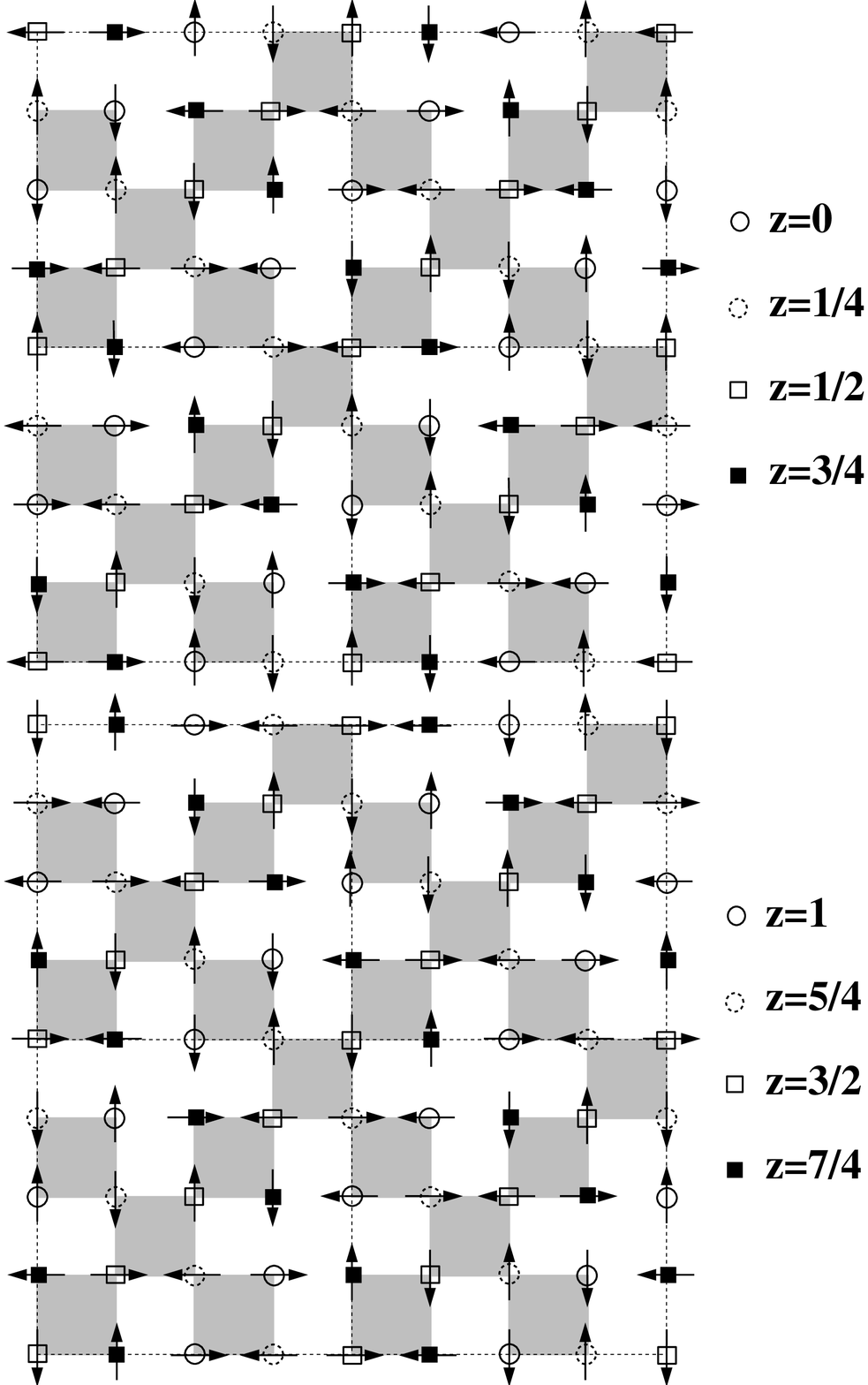}
\caption{
A combination $S^{\{{\bf k}(1,0,\frac{1}{2})\}}$
+ $S^{\{{\bf k}(\frac{1}{2},\frac{1}{2},0)\}}$ for \znco~
which is consistent with all the experimental data explained in the text.
Here $S^{\{{\bf k}(1,0,\frac{1}{2})\}}=\psi^{{\bf k_1} \tau_{11}}
+ \psi^{{\bf k_2} \tau_{12}}
+ i \psi^{{\bf k_1} \tau_{11}^{'}} - i  \psi^{{\bf k_2} \tau_{12}^{'}}
- i \psi^{{\bf k_1} \tau_{21}} - i  \psi^{{\bf k_2} \tau_{22}}
-  \psi^{{\bf k_1} \tau_{21}^{'}} +  \psi^{{\bf k_2} \tau_{22}^{'}}$
with ${\bf k_1}=(1,0,\frac{1}{2})$
and ${\bf k_2}=(\frac{1}{2},0,\bar{1},0)$ (shown in Fig. 5 (c)),
and
$S^{\{{\bf k}(\frac{1}{2},\frac{1}{2},0)\}}=
\psi^{{\bf k_1} \tau_{1}^{'}}+ i \psi^{{\bf k_2} \tau_{1}^{'}}
+ (1+i) \psi^{{\bf k_1} \tau_{2}} + (1-i)  \psi^{{\bf k_2} \tau_{2}}
-i \psi^{{\bf k_1} \tau_{2}^{'}} -  \psi^{{\bf k_2} \tau_{2}^{'}}$
with ${\bf k_1}=(\frac{1}{2},\frac{1}{2},0)$
and ${\bf k_2}=(\bar{\frac{1}{2}},\bar{\frac{1}{2}},0)$ (shown in Fig. 6 (a)).
}
\end{center}
\end{figure}


Now, it is possible that \znco~ has two magnetic domains: one
with ${\bf k}=(1,0,\frac{1}{2})$
and the other with ${\bf k}=(\frac{1}{2},\frac{1}{2},0)$.
However, to explain the neutron powder diffraction data,
the population of the two domains as well as
the ordered moment have to be exactly the same for both domains.
Furthermore, another spinel $\rm ZnFe_2O_4$ also
magnetically orders at low temperatures
and the N\'{e}el state has only a single characteristic wave vector
${\bf k}=(1,0,\frac{1}{2})$.
We believe it is more likely that the two characteristic wavevectors participate
in the ordering of all the spins in \znco.
Then, the resulting spin structure would be
a summation of
$S^{\{{\bf k}(1,0,\frac{1}{2})\}}$
and $S^{\{{\bf k}(\frac{1}{2},\frac{1}{2},0)\}}$.
All Cr$^{3+}$ ions are equivalent in this spinel crystal structure
and are expected to have the same
magnitude, indicating that
$S^{\{{\bf k}(1,0,\frac{1}{2})\}}$
and $S^{\{{\bf k}(\frac{1}{2},\frac{1}{2},0)\}}$
have to be collinear
and orthogonal to each other.
Among the spin structures shown in Fig. 6 and 7,
the only possibility would be
the combination of $S^{\{{\bf k}(1,0,\frac{1}{2})\}}$ shown in Fig. 6 (c)
and $S^{\{{\bf k}(\frac{1}{2},\frac{1}{2},0)\}}$ shown in Fig. 7 (a).
Fig. 8 shows the resulting coplanar and noncollinear spin structure,
in which each tetrahedra has two pairs of antiparallel spins and have
zero net moment.

For the ${\bf k}=(\frac{1}{2},\frac{1}{2},\frac{1}{2})$ domain,
there are many possible spin structures.  From Figure 3 (b)-(d),
we see that the intensities of the ${\bf
k}=(\frac{1}{2},\frac{1}{2},\frac{1}{2})$ reflections relative to
other propagation vectors varies from sample to sample.  These
reflections are strongest for sample 1 shown in Figure 3(b) and it
is only for this sample that it is possible to distinguish between
various models for the spin structure through goodness of fit.
Thus, our discussion will be limited to sample 1.

The basis vectors are given table VI. As the transition is first
order, multiple propagation vectors can contribute to the
ordering.  Many can be ruled out on physical grounds, however, a
large number of possibilities remain. If one visualizes the B
sublattice of the spinel lattice along the $<111>$ direction, one
can decompose it into alternating triangular and kagome layers.
The $\tau_{1}$ configuration, has no spins in the triangular layer
and the "q=0" spin configuration in the kagome layer.  The
$\tau_{2}$ configuration has spins only in the triangular layer
coupled antiferromagnetically.  The $\tau_{3}$ configuration is
ferromagnetic in the kagome plane.  The $\tau_{6}$ configuration
allows for spins in the triangular layer to lie within that plane
at an arbitrary angle.  We also considered linear combinations of
these configurations.

\begin{figure}[!htb]
\begin{center}
\includegraphics[width=3.4in]{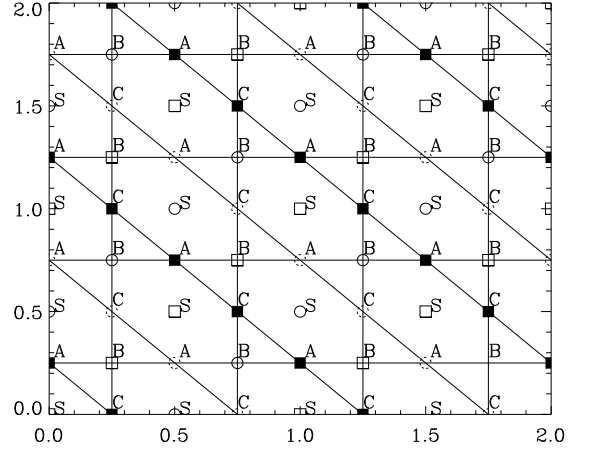}
\caption{
This is the
magnetic structure most consistent with the data for the ${\bf
k}=(\frac{1}{2},\frac{1}{2},\frac{1}{2})$ propagation vector.  The
vertices represent spins.  Spins at vertices A,B, and C lie in the
kagome layer and lie within the plane and have basis vectors of
$(0,1,\bar{1})$, $(\bar{1},0,1)$, and $(1,\bar{1},0)$
respectively. The vertices labeled by S lie in the triangular
layer and the spins point along the body diagonal $(1,1,1)$. The
spins in the triangular layer form an antiferromagnetic pattern,
alternating pattern out/into of the $<1,1,1>$ plane.
}
\end{center}
\end{figure}

In table VII, we show the relative goodness of fits for the
various spin models.  Overall, we found that the model most
consistent with our data was one with the "q=0" configuration in
the kagome layer and an antiferromagnetic configuration within the
triangular layer (see Fig. 9).  Though the difference in the
values of $\chi^2$ is slight, if we examine a nuclear and a
magnetic peak for goodness of fit for two different configurations, $\tau_1 + \tau_2$ and $\tau_1 +\tau_2 +\tau_3$ models, we can see that there is is a real improvement in
the fit for the $\tau_1 + \tau_2$ spin model shown in Fig. 9 (see Fig. 10 and Table VII).  

\begin{figure}[!htb]
\begin{center}
\includegraphics[width=3in]{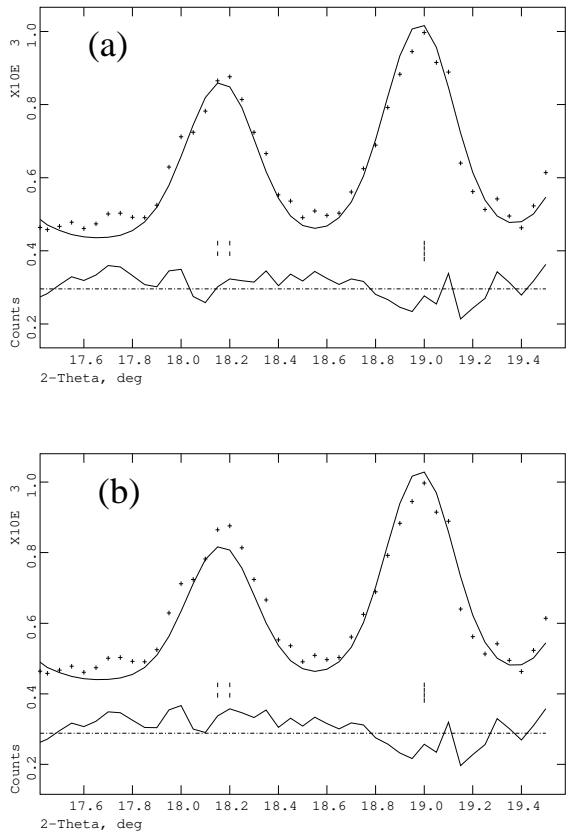}
\caption{
The peak to
the left is the ${\bf
k}=(\bar{\frac{1}{2}},\bar{\frac{1}{2}},\bar{\frac{3}{2}})$
magnetic reflection. The peak to the right is the nuclear $(1,1,1)
$ peak for comparison. (a) the fit to the spin configuration
formed from a linear combination of $\tau_1$ and $\tau_2$.  (b)
fit to the configuration formed formed from a linear combination
of $\tau_1$,$\tau_2$, and $\tau_3$.
}
\end{center}
\end{figure}

\subsection{Summary}

We have solved the magnetic structure of $\rm ZnCr_{2}O_{4}$.  We
have examined single crystals and three polycrystalline samples.
From this, we have found that the system has different domains
formed from different k-vectors.  The relative phase fractions
vary from sample to sample (see Table IX).  The ${\bf k}=(1,0,0)$
is the same collinear structure as that of $ZnV_{2}O_{4}$.  The
${\bf k}=(1,0,\frac{1}{2})$ and ${bf
k}=(\frac{1}{2},\frac{1}{2},0)$ domain has the in plane structure
found in Fig. 8. The structure of the ${\bf
k}=(\frac{1}{2},\frac{1}{2},\frac{1}{2})$ domain is found in Fig.
9.  Since the antiferromagnetic transition is first order,
multiple characteristic wave-vectors are allowed unlike in the
usual case of second order magnetic transitions.  This
multiplicity of domains with different characteristic wave vectors
suggests that even the ordered state of this frustrated magnet
is degenerate.

\section{Magnetic Correlations in $\zcgo$}

In this section, we study how magnetic correlations change with nonmagnetic
doping.

\subsection{First order transition to N\'{e}el state
in $\rm ZnCr_{2}O_4$}

For completeness, we start with the phase transition in pure
\znco\cite{zcoprl}. Fig. 10 shows that in the pure \znco~ long
range antiferromagnetic order (squares in frame (b)) and the local
spin resonance (frame (a)) appear simultaneously in a spectacular
first order phase transition. It also shows that the magnetic
order is accompanied by a cubic to tetragonal lattice distortion
(circles in frame (b)). The tetragonal distortion lifts some of
degeneracy due to geometrical frustration and allows the system to
order magnetically. Furthermore the ordered state pushes spectral
weight in the energy spectrum up to the local spin resonance at
$\hbar\omega\approx 4.5$ meV. It is unusual that a long range
ordered phase can support a local spin resonance.

\begin{figure}[!htb]
\begin{center}
\includegraphics[width=2.6in]{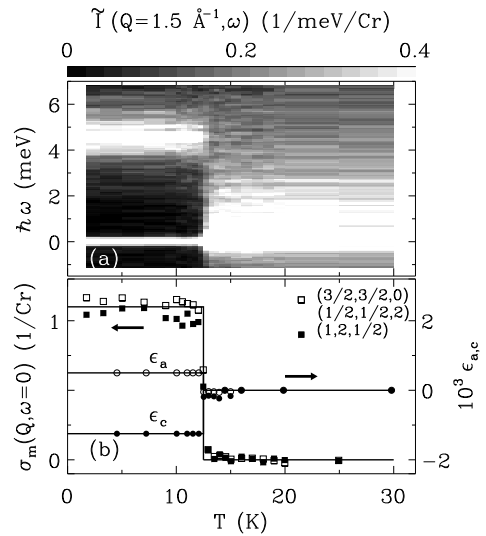}
\caption{
Contour map of inelastic neutron scattering for $Q=1.5$
\AA$^{-1}$. (b) $T-$dependence of magnetic Bragg scattering from a
powder (squares), $\sigma_m=\frac{v_m}{(2\pi)^3}
\int\tilde{I}(Q,\omega)4\pi Q^2dQd\hbar\omega$ where $v_m$ is the
volume per $Cr^{3+}$ ion, and of lattice strain along $\bf a$ and
$\bf c$ (circles) measured by single crystal neutron diffraction.
The figure is reproduced from Ref. \cite{zcoprl}.
}
\end{center}
\end{figure}

\subsection{Second order transition to N\'{e}el state
in $\rm ZnCr_{1.9}Ga_{0.1}O_4$}

The weak nonmagnetic doping in $\rm ZnCr_{1.9}Ga_{0.1}O_4$ does
not change the nature of the low $T$ phase. As shown in Fig. 11,
below $T_N\approx 12.5$ K magnetic long range order (squares in
Fig. 11  (b)), tetragonal distortion (circles in Fig. 11 (b))
occur along with the appearance of the local spin resonance at
$\hbar\omega \approx 4.5$ meV (Fig. 11 (a)). Magnetic peaks in the
doped material are the same as those in \znco, which indicates
that 5\% doping of nonmagnetic Ga$^{3+}$ ions into Cr sites does
not change the spin structure in the ordered phase. However, the
three features appear gradually in a second order fashion, which
is in contrast with the first order phase transition in the pure
\znco~ shown in Fig. 10. Fig. 11 (c) shows that as soon as the
static moment develops FWHM of magnetic peaks becomes
$Q$-resolution limited. This indicates the static correlations are
long range no matter how small the static moment is. We conclude
that the magnetic ordering in the weakly doped material
immediately develops in the entire material rather than in small
magnetic clusters, growing in size grows as $T$ decreases.
\begin{figure}[!htb]
\begin{center}
\includegraphics[width=3.4in]{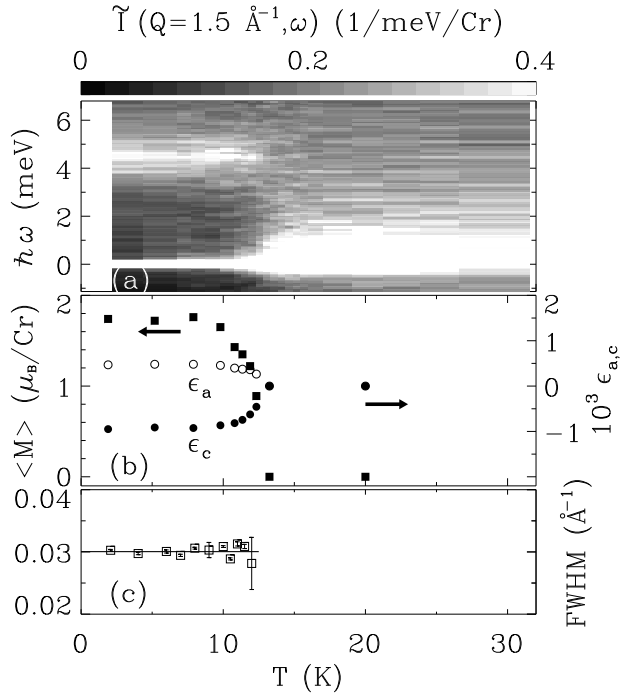}
\caption{
Contour maps of inelastic neutron scattering measured at $Q=1.5$
\AA$^{-1}$ as a function of energy transfer $\hbar\omega$ and $T$.
(b) $T$-dependence of the ordered moment (filled squares) and of
crystal strains (opend and filled circles) obtained by Rietveld
analysis neutron powder diffraction data taken at BT1, NIST at
various $T$s using GSAS (c) $T$-dependence of
full-width-of-Half-Maximum (FWHM) of the magnetic (1/2,1/2,2)
Bragg peak (squares). The line is instrumental angular resolution.
}
\end{center}
\end{figure}

\begin{figure}[!htb]
\begin{center}
\includegraphics[width=2.6in,angle=90]{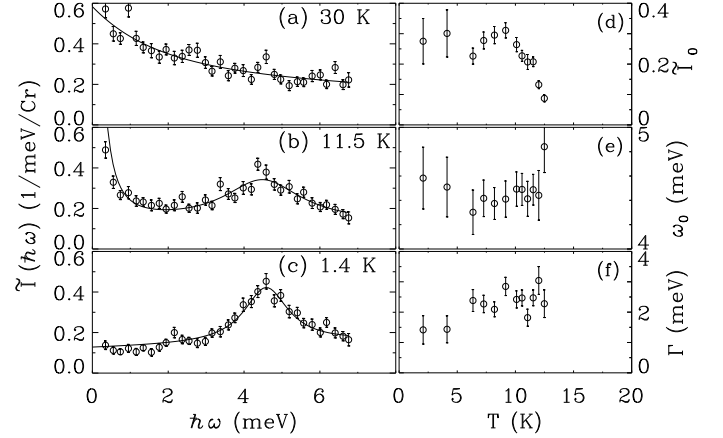}
\caption{
(a)-(c) $\hbar\omega$-dependence of the inelastic magnetic neutron
scattering intensity measured at $Q=1.5 \AA^{-1}$ at three
different $T$s spanning the phase transition. Solid lines are the
fits described in the text. (d) $T$-dependence of integrated
intensity in the unit of (1/meV/Cr), (e) of peak position, and (f)
of full-width-at-half-maximum of the $\hbar\omega \approx 4.5$ meV
excitations. 
}
\end{center}
\end{figure}

Fig. 11 (a) shows that the local spin resonance and the low energy
cooperative paramagnetic spin fluctuations coexist over the
temperature range, 10 K $\leq T \leq T_N = $12.5 K (also see Fig.
12 (b)). For comparison, in the pure \znco~ the Ne\'{e}l phase has
only the linear spin waves below the local spin resonance (see
Fig. 10 (a)). To quantitatively study how the dynamic spin
fluctuations in the weakly doped system evolve with $T$, we have
fit the of $\tilde{I}$ (Q=1.5 \AA$^{-1}$,$\hbar\omega$) in Fig. 11
(a) to two simple non-resonant response functions, each with
single relaxation rate: one  centered at $\hbar\omega=0$ with a
relaxation rate $\Gamma_1$ and the other centered at
$\hbar\omega_0 \approx 4.5$ meV with a relaxation rate $\Gamma$.
\begin{equation}
\tilde{I} (\hbar\omega) = \frac{\tilde{I_1}\cdot
(\frac{\Gamma_1}{2})^2}{ (\hbar\omega)^2
+ (\frac{\Gamma_1}{2})^2}
+ \frac{\tilde{I}_0\cdot (\frac{\Gamma}{2})^2}{ (\hbar\omega-\hbar\omega_0)^2
+ (\frac{\Gamma}{2})^2}.
\end{equation}
The first term is to account for the quasi-elastic scattering that
exists at $T > 10$ K. Since we did not have data above 30 K, it is
difficult to extract meaningful information on the low energy
excitations due to cooperative paramagnetism. Here we focuse on
the local spin resonance. Fig. 12 (d) - (f) show the results of
the fits. The peak position, $\hbar\omega \approx 12.5$ meV, (Fig.
12 (e)) is $T$-independent below $T_N$ within experimental
accuracy. The relaxation rate of the local resonance decreases as
$T$ decease to $\Gamma = 1.4(1)$ meV at 1.4 K. For comparison, in
pure \znco~ $\Gamma = 1.5(1)$ meV for all $T<T_N$. The strength of
the local resonance, $\tilde{I}_0$ (Fig. 12 (d)), develops as
proportional to the staggered magnetization, $<M>$, shown is Fig.
11 (b). This suggests that the static spin component is necessary
to support the local spin resonance.

\begin{figure}[!htb]
\begin{center}
\includegraphics[width=2.6in,angle=90]{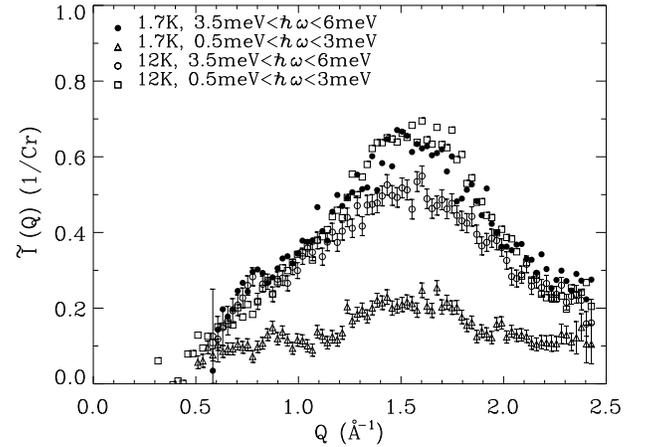}
\caption{
Q-dependence of the inelastic magnetic neutron scattering
intensity at 1.7 K and 12 K which is integrated over different
energies.
}
\end{center}
\end{figure}

Fig. 13 shows the spatial correlations of the fluctuating moments
with different energies. The low energy lying excitations at 12 K
(squares) and the local resonance at 1.7 K (filled circles) and 12
K (open circles) have almost identical $Q$-dependence with a Half
Width at Half Maximum $\kappa = 0.50(5)$ \AA$^{-1} = 0.67(6) a^*$.
Even though their characteristic energies are different, the
structure factor associated with the spin fluctuations have the
same wave vector dependence. The excitations for $\hbar\omega <$ 3
meV at 1.7 K (triangles) also have a broad peak centered at
$Q=$1.5 \AA$^{-1}$.

\subsection{Spin freezing
in $\rm ZnCr_{1.4}Ga_{0.6}O_4$}

\begin{figure}[!htb]
\begin{center}
\includegraphics[width=2.6in,angle=90]{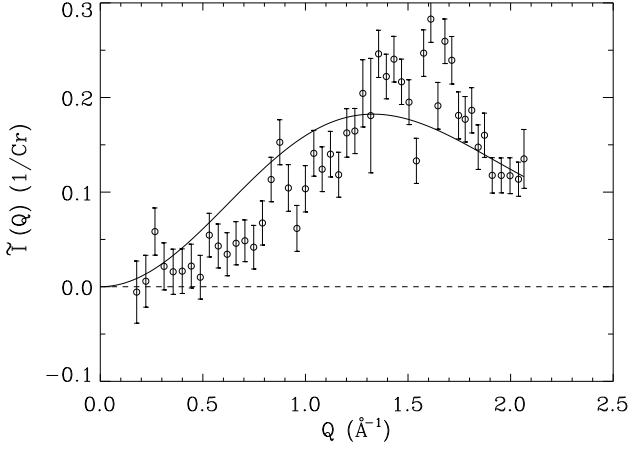}
\caption{
Q-dependence of elastic magnetic neutron scattering intensity
measured with energy window of $|\hbar\omega | < 0.05$ meV.
}
\end{center}
\end{figure}

In this section, we study spin correlations in $\rm
ZnCr_{1.4}Ga_{0.6}O_4$ which exhibits spin-glass-like behaviors in
bulk susceptibility measurements\cite{fiorani84}. Fig. 14 shows
elastic magnetic scattering intensities measured at 1.4 K. High
temperature background was measured at 20 K and subtracted. Unlike
in the pure \znco~ and the weakly doped \gapone, this system does
not have magnetic Bragg peaks but a broad peak centered at a
finite wave vector $Q=1.5$ \AA$^{-1}$ with $\kappa=0.48(5)$
\AA$^{-1}$. This indicates that the 30\% nonmagnetic doping
destroys the magnetic long range order and reduces the correlation
length down to distance between nearest neighboring Cr ions.
$\tilde{I} (Q)$ going to zero as Q approaches zero indicates that
the antiferromagnetic constraints are still satisfied in the
heavily doped sample. The solid line is the fit to the
powder-averaged magnetic neutron scattering intensity for an
isolated spin dimer\cite{furr79,scgoprl},
\begin{equation}
 \tilde{I}(Q)\propto |F(Q)|^2\frac{1-{\rm sin}Qr_0}{Qr_0}
\end{equation}
where the distance between nearest neighboring Cr$^{3+}$ ions
$r_0=2.939$~\AA. The spin pair model produces a broader peak than
the experimental data. Instead, the $Q$-dependence is almost
identical to that of the fluctuating spins in pure \znco~ and
weakly diluted $\rm ZnCr_{1.9}Ga_{0.1}O_4$ (see Fig. 13). This
suggests that the same local spin objects involving more than
isolated dimers are responsible for the broad $Q$-dependence in
those materials even though the energy for the correlations
changes with doping of the nonmagnetic ions. We can estimate the
average frozen moment from the elastic neutron scattering data
\begin{equation}
|<M>|^2 \approx \frac{\frac{3}{2} g^2\int_{0.2\AA^{-1}}^{2.1\AA^{-1}}
(\tilde{I}(Q)/|F(Q)|^2)Q^2dQ}{
\int_{0.2\AA^{-1}}^{2.1\AA^{-1}} Q^2/dQ}.
\end{equation}
Integrating the difference data over $Q$ yields $|<M>|^2 = 0.24(4) \mu_B^2$/Cr,
in other words, $|<M>|= g<S> \mu_B = 0.49(4)\mu_B$/Cr.
This quantity is substantially less than the N\'{e}el value
$|<M>|=gS\mu_B=3\mu_B$/Cr and also much less than those values
of \znco~ and \gapone.

\begin{figure}[!htb]
\begin{center}
\includegraphics[width=2.6in,angle=90]{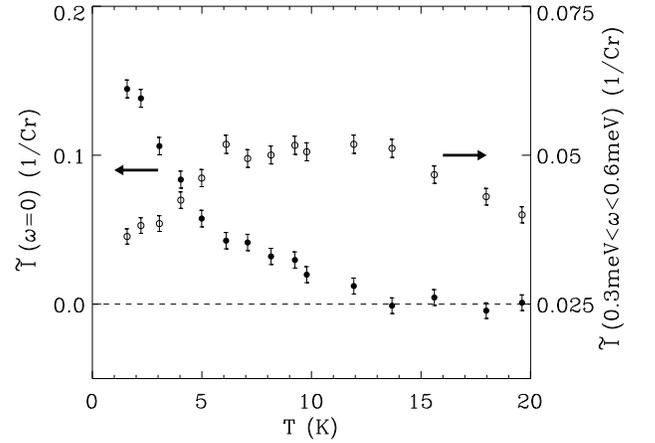}
\caption{
$T$-dependence of elastic magnetic neutron scattering intensity
interaged over $|\hbar\omega|<$ 0.05 meV and inelastic magnetic
neutron scattering intensity integrated over 0.3 meV
$<\hbar\omega<$ 0.6 meV. Both data were measured at $Q=1.5$
\AA$^{-1}$. 
}
\end{center}
\end{figure}

Fig. 15 shows elastic and inelastic neutron scattering intensities
measured at $Q=1.5$ \AA$^{-1}$ as a function of temperature. The
onset of elastic magnetic scattering at around 10 K signals the
development of magnetic correlations on a time scale, $\tau >
2\hbar/\Delta E = 0.013$ ns set by the energy resolution of the
instrument. Energy integrated inelastic scattering data over 0.3
meV $< \hbar\omega <$ 0.6 meV show a broad maximum at around 10 K
indicating the critical scattering at the phase transition. Bulk
susceptibility data with a maximum at a lower temperature
$T_f=1.8$ K\cite{fiorani84} show that this apparent critical
temperature is not unique but decreases with the energy scale of
the measurement. Such behavior, which is characteristic of spin
glasses, indicates that a precipitous softening of the magnetic
fluctuation spectrum takes place upon lowing the temperature,
leading to anomalies when the lowest energy scale of the system
falls below the characteristic energy scale of the measurement.

\section{Discussions and Summary}

\znco~ is so far the best realization of an antiferromagnet on the
magnetic lattice of corner-sharing tetrahedra with uniform nearest
neighbor interactions and without any site-disorder. Upon cooling,
this system is heading toward spin-liquid state with the signature
of almost linear spin relaxation rate. At low temperature
$T<T_N=12.5$ K , the system undergoes a cubic-to-tetragonal
distortion to settle into a N\'{e}el phase with a local spin
resonance\cite{zcoprl}. The three features, tetragonal distortion,
long range order, and the local spin resonance, occurs abruptly in
a first order fashion. The spins in the N\'{e}el phase have
reduced staggered magnetization, $\frac{<S>}{S}<1$ (see the inset
of Fig. 2) due to geometrical frustration. Weak 5\% nonmagnetic
doping into the magnetic lattice further suppresses the staggered
magnetization but does not destroy the N\'{e}el phase at low
temperatures. The phase transition from cooperative paramagnetic
phase to N\'{e}el phase, however, occurs gradually upon cooling in
a second order fashion. The cubic to tetragonal lattice distortion
also follows the development of the magnetic phase transition.
This supports that the phase transition is magnetically driven. It
is understandable that the nature of the low temperature phase
does not change with 5\% doping because for 5\% dilution in the
magnetic lattice, the majority of tetrahedra have all 4 spins
(81\% of tetrahedra have all 4 spins and 17\% have 3
spins\cite{moeber99}). For 30\% dilution (1-x=0.7), 24\% of
tetrahedra have 4 spins, 41\% have 3 spins and 27\% have 2 spins.
Even though it is still above the percolation threshold
($1-x_c=0.39$), the long range correlations are destroyed and
replaced with short range ocrrelations. Despite these differences,
all three materials contain spin correlations with a common broad
$Q$ dependence even though the energetics of the local
correlations change with the occupance of the magnetic lattice and
the existence of long range order. This indicates that the local
spin object responsible for the common $Q$ dependence is robust
against strong disorder. This finding may explain why bulk
properties in geometrically frustrated magnets are robust against
dilution. In SCGO(x), where the magnetic entity relevant to
geometrical frustration can be viewed as quasi-two dimensional
(111)-slabs of corner-sharing tetrahedra\cite{scgoprl}, the bulk
susceptibility shows field hysteresis and the nonlinear
susceptibility diverges, typical of spin glasses, the specific
heat $C(T)$ is proportional to $T^2$ as in an ordinary
two-dimensional antiferromagnet\cite{apr92}. These bulk behaviors
are very robust against magnetic dilution.\cite{apr92, schiffer}
Our finding indicates that the fundamental spin degree of freedom
in \zcgo~ is the hexagonal loop of {\it antiferromagnetic}
spins observed in the pure ZnCr$_2$O$_4$.\cite{zconature} The low energy physics is governed by the excitations of
the local spin degree of freedom and therefore is robust to
dilution.

In summary, we have determined the spin structure of
the N\'{e}el phase in \zcgo~ which would provide a starting
point for a theory for this system. We have studied, using neutron scattering,
how nonmagnetic doping
changes the first order magnetoelastic phase transition
in pure \znco~
into the second order spin-glass-like phase transition.
We have found that a broad $Q$ dependence is robust against dilution,
suggesting that such local spin correlations both in N\'{e}el phase
and in short range ordered phase is intrinsic to the geometrically
frustrated magnets and distinguishes these systems
 from the ordinary spin glasses.

\acknowledgements

We thank C. Broholm for helpful discussions. 
S.H.L is supported by the U.S. DOE through DE-FG02-07ER45384.
The NSF supported work at Rutgers through DMR-9802513 and work at SPINS through DMR-0454672.

\begin{widetext}

\begin{table}[htb]
\caption{Measured ratios of SF to NSF scattering 
intensities for three magnetic
reflections that belong to ${\bf k}=(1,0,\frac{1}{2})$.
}
\begin{tabular}{ccc}  
$(h,l,k)$  & $(\sigma_{SF}/\sigma_{NSF})_{obs}$  
& $1-\frac{h^2}{h^2+l^2}$ \\\hline
(1,0.5,0) & 0.4(1) & 0.2 \\
(1,1.5,0) & 0.7(1) & 0.6923 \\
(1,2.5,0) & 0.83(5) & 0.862 \\
\end{tabular}
\end{table}

\begin{table}[htb]
\caption{The basis functions of the irreducible representation
of group $O_h^7$ for the two-arm star of the wave vector 
${\bf k}=(\frac{1}{2},\frac{1}{2},0)$. Here ${\bf k}$ is in terms of the cubic notation 
and ${\bf k}=\frac{1}{4}(2,1,1)$ in Kovalev's notation\cite{kova}.
Here ${\bf k}_1={\bf k}$ and ${\bf k}_2=-{\bf k}$. The notation of representations, such as $\tau_1$,
$\tau_1^{'}$ and so on, followed the Kovalev's notation.
$\psi^{{\bf k}\tau}_i (i=1,2,3,4)$ represent the basis functions for the 
spins located at (0.5,0.5,0.5), (0.5,0.75,0.75),
(0.75,0.5,0.75), and (0.75,0.75,0.5) in the cubic notation, respectively. 
This table was taken from Table  22 in page 131 of Ref. \cite{izyumov}.
}
\begin{tabular}{cccccc}  
Rep. & Arms 
& $\psi^{{\bf k}\tau}_1$ 
& $\psi^{{\bf k}\tau}_2$ 
& $\psi^{{\bf k}\tau}_3$ 
& $\psi^{{\bf k}\tau}_4$ 
\\\hline
$\tau_1$ & $k_1$ & $(1,\bar{1},0)$ & (0,0,0) & (0,0,0) & $i (\bar{1},1,0)$ \\
  & $k_2$ & $(1,\bar{1},0)$ & (0,0,0) & (0,0,0) & $-i (\bar{1},1,0)$ \\
$\tau_1^{'}$ & $k_1$ & (0,0,0) & $(1,1,0)$ & $(\bar{1},\bar{1},0)$ & $(0,0,0)$ \\
  & $k_2$ & (0,0,0) & $-i (1,1,0)$ & $-i (\bar{1},\bar{1},0)$ & $(0,0,0)$ \\
$\tau_2$ & $k_1$ & $(1,1,0)$ & (0,0,0) & (0,0,0) & $i (1,1,0)$ \\
  & $k_2$ & $(1,1,0)$ & (0,0,0) & (0,0,0) & $-i (1,1,0)$ \\
$\tau_2^{'}$ & $k_1$ & (0,0,0) & $(1,1,0)$ & $(1,1,0)$ & $(0,0,0)$ \\
  & $k_2$ & (0,0,0) & $-i (1,1,0)$ & $-i (1,1,0)$ & $(0,0,0)$ \\
$\tau_2^{''}$ & $k_1$ & $(0,0,1)$ & (0,0,0) & (0,0,0) & $i (0,0,\bar{1})$ \\
  & $k_2$ & $(0,0,1)$ & (0,0,0) & (0,0,0) & $-i (0,0,\bar{1})$\\
$\tau_3$ & $k_1$ & $(1,1,0)$ & (0,0,0) & (0,0,0) & $i (\bar{1},\bar{1},0)$ \\
  & $k_2$ & $(1,1,0)$ & (0,0,0) & (0,0,0) & $-i (\bar{1},\bar{1},0)$ \\
$\tau_3^{'}$ & $k_1$ & (0,0,0) & $(1,\bar{1},0)$ & $(\bar{1},1,0)$ & $(0,0,0)$ \\
  & $k_2$ & (0,0,0) & $-i (1,\bar{1},0)$ & $-i (\bar{1},1,0)$ & $(0,0,0)$ \\
$\tau_3^{''}$ & $k_1$ & $(0,0,1)$ & (0,0,0) & (0,0,0) & $i (0,0,1)$ \\
  & $k_2$ & $(0,0,1)$ & (0,0,0) & (0,0,0) & $-i (0,0,1)$\\
$\tau_3^{'''}$ & $k_1$ &  (0,0,0) & $(0,0,1)$ & $(0,0,1)$ &(0,0,0) \\
  & $k_2$ &  (0,0,0) & $-i (0,0,1)$ & $-i (0,0,1)$ &(0,0,0) \\ 
$\tau_4$ & $k_1$ & $(1,\bar{1},0)$ & (0,0,0) & (0,0,0) & $i (1,\bar{1},0)$ \\
  & $k_2$ & $(1,\bar{1},0)$ & (0,0,0) & (0,0,0) & $-i (1,\bar{1},0)$ \\
$\tau_4^{'}$ & $k_1$ & (0,0,0) & $(1,\bar{1},0)$ & $(1,\bar{1},0)$ & $(0,0,0)$ \\
  & $k_2$ & (0,0,0) & $-i (1,\bar{1},0)$ & $-i (1,\bar{1},0)$ & $(0,0,0)$ \\
$\tau_4^{''}$ & $k_1$ &  (0,0,0) & $(0,0,1)$ & $(0,0,\bar{1})$ &(0,0,0) \\
  & $k_2$ &  (0,0,0) & $-i (0,0,1)$ & $-i (0,0,\bar{1})$ &(0,0,0) 
\end{tabular}
\end{table}

\begin{table}[htb]
\caption{The basis functions of the irreducible representation
of group $Fd\bar{3}m (O_h^7)$ for the two-arm star of the wave vector 
${\bf k}=(1,0,\frac{1}{2})$. Here ${\bf k}$ is in terms of the cubic notation 
and ${\bf k}=\frac{1}{4}(1,1,0)+\frac{1}{2}(0,1,1)$ in Kovalev's notation
using primitive reciprocal unit vectors \cite{kova}.
Here ${\bf k}_1={\bf k}$ and ${\bf k}_2=-{\bf k}$. The notation of representations, such as $\tau_1$,
$\tau_1^{'}$ and so on, followed the Kovalev's notation.
$\psi^{{\bf k}\tau}_i (i=1,2,3,4)$ 
represent the basis functions for four 
spins in the primitive cell located at (0.5,0.5,0.5), (0.5,0.75,0.75),
(0.75,0.5,0.75), and (0.75,0.75,0.5) in the cubic notation, respectively. 
}
\begin{tabular}{cccccc}  
Rep. & Arms 
& $\psi^{{\bf k}\tau}_1$ 
& $\psi^{{\bf k}\tau}_2$ 
& $\psi^{{\bf k}\tau}_3$ 
& $\psi^{{\bf k}\tau}_4$ 
\\\hline
$\tau_{11}$ & $k_1$ & $(1,0,0)$ & $-i (0,1,0)$ & $-i (0,1,0)$ & $(1,0,0)$ \\
  & $k_2$ & $(\bar{1},0,0)$ & $(0,1,0)$ & $(0,\bar{1},0)$ & $(1,0,0)$ \\
$\tau_{12}$ & $k_1$ & $(\bar{1},0,0)$ & $(0,1,0)$ & $(0,\bar{1},0)$ & $(1,0,0)$ \\
  & $k_2$ & $(1,0,0)$ & $i (0,1,0)$ & $i (0,1,0)$ & $(1,0,0)$ \\
$\tau_{11}^{'}$ & $k_1$ & $(0,1,0)$ & $i (1,0,0)$ & $i (1,0,0)$ & $(0,1,0)$ \\
  & $k_2$ & $(0,\bar{1},0)$ & $(\bar{1},0,0)$ & $(1,0,0)$ & $(0,1,0)$ \\
$\tau_{12}^{'}$ & $k_1$ & $(0,\bar{1},0)$ & $(\bar{1},0,0)$ & $(1,0,0)$ & $(0,1,0)$ \\
  & $k_2$ & $(0,1,0)$ & $-i (1,0,0)$ & $-i (1,0,0)$ & $(0,1,0)$ \\
$\tau_{11}^{''}$ & $k_1$ & $(0,0,1)$ & $-i (0,0,1)$ & $i (0,0,1)$ & $(0,0,\bar{1})$ \\
  & $k_2$ & $(0,0,\bar{1})$ & $(0,0,1)$ & $(0,0,1)$ & $(0,0,\bar{1})$ \\
$\tau_{12}^{''}$ & $k_1$ & $(0,0,\bar{1})$ & $(0,0,1)$ & $(0,0,1)$ & $(0,0,\bar{1})$ \\
  & $k_2$ & $(0,0,1)$ & $i (0,0,1)$ & $-i (0,0,1)$ & $(0,0,\bar{1})$ \\
$\tau_{21}$ & $k_1$ & $(1,0,0)$ & $i (0,1,0)$ & $i (0,1,0)$ & $(1,0,0)$ \\
  & $k_2$ & $(\bar{1},0,0)$ & $(0,\bar{1},0)$ & $(0,1,0)$ & $(1,0,0)$ \\
$\tau_{22}$ & $k_1$ & $(1,0,0)$ & $(0,1,0)$ & $(0,\bar{1},0)$ & $(\bar{1},0,0)$ \\
  & $k_2$ & $(\bar{1},0,0)$ & $i (0,1,0)$ & $i (0,1,0)$ & $(\bar{1},0,0)$ \\
$\tau_{21}^{'}$ & $k_1$ & $(0,1,0)$ & $i (\bar{1},0,0)$ & $i (\bar{1},0,0)$ & $(0,1,0)$ \\
  & $k_2$ & $(0,\bar{1},0)$ & $(1,0,0)$ & $(\bar{1},0,0)$ & $(0,1,0)$ \\
$\tau_{22}^{'}$ & $k_1$ & $(0,1,0)$ & $(\bar{1},0,0)$ & $(1,0,0)$ & $(0,\bar{1},0)$ \\
  & $k_2$ & $(0,\bar{1},0)$ & $-i (1,0,0)$ & $-i (1,0,0)$ & $(0,\bar{1},0)$ \\
$\tau_{21}^{''}$ & $k_1$ & $(0,0,1)$ & $i (0,0,1)$ & $-i (0,0,1)$ & $(0,0,\bar{1})$ \\
  & $k_2$ & $(0,0,\bar{1})$ & $(0,0,\bar{1})$ & $(0,0,\bar{1})$ & $(0,0,\bar{1})$ \\
$\tau_{22}^{''}$ & $k_1$ & $(0,0,1)$ & $(0,0,1)$ & $(0,0,1)$ & $(0,0,1)$ \\
  & $k_2$ & $(0,0,\bar{1})$ & $i (0,0,1)$ & $-i (0,0,1)$ & $(0,0,1)$ 
\end{tabular}
\end{table}

\begin{table}[htb]
\caption{Superpositions of two irreducible representations,
$C_1 \psi^{{\bf k}_1\tau}+C_2 \psi^{{\bf k}_2\tau}$, which give real spins
to the atoms
for ${\bf k}_1=(\frac{1}{2},\frac{1}{2},0)$ and ${\bf k}_2=-{\bf k}_1$.
$1(1,\bar{1},0)$ represents the spin at site 1 is along $(1,\bar{1},0)$.
Positions of the 1 to 16 sites are shown in Fig. 6 (a).
}
\begin{tabular}{ccccc}  
\multicolumn{1}{c}{Superposition} 
& \multicolumn{4}{c}{Nonzero spins in a chemical unit cell} \\\hline
$k_1 \tau_1 + k_2 \tau_1$
& 1$(1,\bar{1},0)$ 
& 8$(\bar{1},1,0)$ 
& 9$(\bar{1},1,0)$ 
& 16$(\bar{1},1,0)$ \\
$-i k_1 \tau_1 + i k_2 \tau_1$ 
& 4$(\bar{1},1,0)$ 
& 5$(\bar{1},1,0)$ 
& 12$(1,\bar{1},0)$ 
& 13$(\bar{1},1,0)$ \\
$k_1 \tau_1^{'} + i k_2 \tau_1^{'}$
& 2$(1,1,0)$ 
& 3$(\bar{1},\bar{1},0)$ 
& 10$(\bar{1},\bar{1},0)$ 
& 11$(1,1,0)$ \\ 
$i k_1 \tau_1^{'} + k_2 \tau_1^{'}$
& 6$(1,1,0)$ 
& 7$(\bar{1},\bar{1},0)$ 
& 14$(1,1,0)$  
& 15$(\bar{1},\bar{1},0)$ \\
$k_1 \tau_2 + k_2 \tau_2$
& 1$(1,1,0)$ 
& 8$(1,1,0)$ 
& 9$(\bar{1},\bar{1},0)$ 
& 16$(1,1,0)$ \\
$-i k_1 \tau_2 + i k_2 \tau_2$
& 4$(1,1,0)$ 
& 5$(\bar{1},\bar{1},0)$ 
& 12$(\bar{1},\bar{1},0)$ 
& 13$(\bar{1},\bar{1},0)$ \\
$k_1 \tau_2^{'} + i k_2 \tau_2^{'}$
& 2$(1,1,0)$ 
& 3$(1,1,0)$ 
& 10$(\bar{1},\bar{1},0)$ 
& 11$(\bar{1},\bar{1},0)$ \\
$i k_1 \tau_2^{'} + k_2 \tau_2^{'}$
& 6$(1,1,0)$ 
& 7$(1,1,0)$ 
& 14$(1,1,0)$ 
& 15$(1,1,0)$ \\
$k_1 \tau_2^{''} + k_2 \tau_2^{''}$
& 1$(0,0,1)$ 
& 8$(0,0,\bar{1})$ 
& 9$(0,0,\bar{1})$ 
& 16$(0,0,\bar{1})$ \\
$-i k_1 \tau_2^{''} + i k_2 \tau_2^{''}$
& 4$(0,0,\bar{1})$ 
& 5$(0,0,\bar{1})$ 
& 12$(0,0,1)$ 
& 13$(0,0,\bar{1})$ \\
$k_1 \tau_3 + k_2 \tau_3$
& 1$(1,1,0)$ 
& 8$(\bar{1},\bar{1},0)$ 
& 9$(\bar{1},\bar{1},0)$ 
& 16$(\bar{1},\bar{1},0)$ \\
$-i k_1 \tau_3 + i k_2 \tau_3$
& 4$(\bar{1},\bar{1},0)$ 
& 5$(\bar{1},\bar{1},0)$ 
& 12$(1,1,0)$ 
& 13$(\bar{1},\bar{1},0)$ \\
$k_1 \tau_3^{'} + i k_2 \tau_3^{'}$
& 2$(1,\bar{1},0)$ 
& 3$(\bar{1},1,0)$ 
& 10$(\bar{1},1,0)$ 
& 11$(1,\bar{1},0)$ \\
$i k_1 \tau_3^{'} + k_2 \tau_3^{'}$
& 6$(1,\bar{1},0)$ 
& 7$(\bar{1},1,0)$ 
& 14$(1,\bar{1},0)$ 
& 15$(\bar{1},1,0)$ \\
$k_1 \tau_3^{''} +k_2 \tau_3^{''}$
& 1$(0,0,1)$ 
& 8$(0,0,1)$ 
& 9$(0,0,\bar{1})$ 
& 16$(0,0,1)$ \\
$-i k_1 \tau_3^{''} +i k_2 \tau_3^{''}$
& 4$(0,0,1)$ 
& 5$(0,0,\bar{1})$ 
& 12$(0,0,\bar{1})$ 
& 13$(0,0,\bar{1})$ \\
$k_1 \tau_3^{'''} +i k_2 \tau_3^{'''}$
& 2$(0,0,1)$ 
& 3$(0,0,1)$ 
& 10$(0,0,\bar{1})$ 
& 11$(0,0,\bar{1})$ \\
$i k_1 \tau_3^{'''} +  k_2 \tau_3^{'''}$
& 6$(0,0,1)$ 
& 7$(0,0,1)$ 
& 14$(0,0,1)$ 
& 15$(0,0,1)$ \\
$k_1 \tau_4 + k_2 \tau_4$
& 1$(1,\bar{1},0)$ 
& 8$(1,\bar{1},0)$ 
& 9$(\bar{1},1,0)$ 
& 16$(1,\bar{1},0)$ \\
$-i k_1 \tau_4 + i k_2 \tau_4$
& 4$(1,\bar{1},0)$ 
& 5$(\bar{1},1,0)$ 
& 12$(\bar{1},1,0)$ 
& 13$(\bar{1},1,0)$ \\
$k_1 \tau_4^{'} + i k_2 \tau_4^{'}$
& 2$(1,\bar{1},0)$ 
& 3$(1,\bar{1},0)$ 
& 10$(\bar{1},1,0)$ 
& 11$(\bar{1},1,0)$ \\
$i k_1 \tau_4^{'} + k_2 \tau_4^{'}$
& 6$(1,\bar{1},0)$ 
& 7$(1,\bar{1},0)$ 
& 14$(1,\bar{1},0)$ 
& 15$(1,\bar{1},0)$ \\
$k_1 \tau_4^{''} + i k_2 \tau_4^{''}$
& 2$(0,0,1)$
& 3$(0,0,\bar{1})$ 
& 10$(0,0,\bar{1})$ 
& 11$(0,0,1)$\\ 
$i k_1 \tau_4^{''} + k_2 \tau_4^{''}$
& 6$(0,0,1)$
& 7$(0,0,\bar{1})$ 
& 14$(0,0,1)$
& 15$(0,0,\bar{1})$ 
\end{tabular}
\end{table}

\begin{table}[htb]
\caption{Superpositions of two irreducible representations,
$C_1 \psi^{{\bf k}_1\tau}+C_2 \psi^{{\bf k}_2\tau}$, which give real spins
to the atoms
for ${\bf k}_1=(1,0,\frac{1}{2})$ and ${\bf k}_2=-{\bf k}_1$.
}
\begin{tabular}{ccccccccc}  
\multicolumn{1}{c}{Superposition} 
& \multicolumn{8}{c}{Nonzero spins in a chemical unit cell} \\\hline
$k_1 \tau_{11} + k_2 \tau_{12}$   
& 1$(1,0,0)$ 
& 4$(1,0,0)$ 
& 6$(0,1,0)$ 
& 7$(0,1,0)$ 
& 9$(\bar{1},0,0)$ 
& 12$(\bar{1},0,0)$ 
& 14$(0,\bar{1},0)$ 
& 15$(0,\bar{1},0)$  \\
$-i k_1 \tau_{11} + i k_2 \tau_{12}$   
& 2$(0,\bar{1},0)$ 
& 3$(0,\bar{1},0)$ 
& 5$(1,0,0)$ 
& 8$(1,0,0)$ 
& 10$(0,1,0)$ 
& 11$(0,1,0)$ 
& 13$(\bar{1},0,0)$ 
& 16$(\bar{1},0,0)$ \\
$k_1 \tau_{12} + k_2 \tau_{11}$   
& 1$(\bar{1},0,0)$ 
& 2$(0,1,0)$ 
& 3$(0,\bar{1},0)$ 
& 4$(1,0,0)$ 
& 9$(1,0,0)$ 
& 10$(0,\bar{1},0)$ 
& 11$(0,1,0)$ 
& 12$(\bar{1},0,0)$ \\
$-i k_1 \tau_{12} + i k_2 \tau_{11}$   
& 5$(\bar{1},0,0)$ 
& 6$(0,1,0)$ 
& 7$(0,\bar{1},0)$ 
& 8$(1,0,0)$ 
& 13$(1,0,0)$ 
& 14$(0,\bar{1},0)$ 
& 15$(0,1,0)$ 
& 16$(\bar{1},0,0)$ \\
$k_1 \tau_{11}^{'} + k_2 \tau_{12}^{'}$   
& 1$(0,1,0)$ 
& 4$(0,1,0)$ 
& 6$(\bar{1},0,0)$ 
& 7$(\bar{1},0,0)$ 
& 9$(0,\bar{1},0)$ 
& 12$(0,\bar{1},0)$ 
& 14$(1,0,0)$ 
& 15$(1,0,0)$ \\
$-i k_1 \tau_{11}^{'} + i k_2 \tau_{12}^{'}$   
& 2$(1,0,0)$ 
& 3$(1,0,0)$ 
& 5$(0,1,0)$ 
& 8$(0,1,0)$ 
& 10$(\bar{1},0,0)$ 
& 11$(\bar{1},0,0)$ 
& 13$(0,\bar{1},0)$ 
& 16$(0,\bar{1},0)$ \\
$k_1 \tau_{12}^{'} + k_2 \tau_{11}^{'}$   
& 1$(0,\bar{1},0)$ 
& 2$(\bar{1},0,0)$ 
& 3$(1,0,0)$ 
& 4$(0,1,0)$ 
& 9$(0,1,0)$ 
& 10$(1,0,0)$ 
& 11$(\bar{1},0,0)$ 
& 12$(0,\bar{1},0)$ \\
$-i k_1 \tau_{12}^{'} + i k_2 \tau_{11}^{'}$   
& 5$(0,\bar{1},0)$ 
& 6$(\bar{1},0,0)$ 
& 7$(1,0,0)$ 
& 8$(0,1,0)$ 
& 13$(0,1,0)$ 
& 14$(1,0,0)$ 
& 15$(\bar{1},0,0)$ 
& 16$(0,\bar{1},0)$ \\
$k_1 \tau_{11}^{''} + k_2 \tau_{12}^{''}$   
& 1$(0,0,1)$ 
& 4$(0,0,\bar{1})$ 
& 6$(0,0,1)$ 
& 7$(0,0,\bar{1})$ 
& 9$(0,0,\bar{1})$ 
& 12$(0,0,1)$ 
& 14$(0,0,\bar{1})$ 
& 15$(0,0,1)$ \\
$-i k_1 \tau_{11}^{''} + i k_2 \tau_{12}^{''}$   
& 2$(0,0,\bar{1})$ 
& 3$(0,0,1)$ 
& 5$(0,0,1)$ 
& 8$(0,0,\bar{1})$ 
& 10$(0,0,1)$ 
& 11$(0,0,\bar{1})$ 
& 13$(0,0,\bar{1})$ 
& 16$(0,0,1)$ \\
$k_1 \tau_{12}^{''} + k_2 \tau_{11}^{''}$   
& 1$(0,0,\bar{1})$ 
& 2$(0,0,1)$ 
& 3$(0,0,1)$ 
& 4$(0,0,\bar{1})$ 
& 9$(0,0,1)$ 
& 10$(0,0,\bar{1})$ 
& 11$(0,0,\bar{1})$ 
& 12$(0,0,1)$ \\
$-i k_1 \tau_{12}^{''} + i k_2 \tau_{11}^{''}$   
& 5$(0,0,\bar{1})$ 
& 6$(0,0,1)$ 
& 7$(0,0,1)$ 
& 8$(0,0,\bar{1})$ 
& 13$(0,0,1)$ 
& 14$(0,0,\bar{1})$ 
& 15$(0,0,\bar{1})$ 
& 16$(0,0,1)$ \\
$k_1 \tau_{21} - k_2 \tau_{22}$   
& 1$(1,0,0)$ 
& 4$(1,0,0)$ 
& 6$(0,\bar{1},0)$ 
& 7$(0,\bar{1},0)$ 
& 9$(\bar{1},0,0)$ 
& 12$(\bar{1},0,0)$ 
& 14$(0,1,0)$ 
& 15$(0,1,0)$ \\
$i k_1 \tau_{21} + i k_2 \tau_{22}$   
& 2$(0,\bar{1},0)$ 
& 3$(0,\bar{1},0)$ 
& 5$(\bar{1},0,0)$ 
& 8$(\bar{1},0,0)$ 
& 10$(0,1,0)$ 
& 11$(0,1,0)$  
& 13$(1,0,0)$ 
& 16$(1,0,0)$ \\
$- k_1 \tau_{22} + k_2 \tau_{21}$   
& 1$(\bar{1},0,0)$ 
& 2$(0,\bar{1},0)$ 
& 3$(0,1,0)$ 
& 4$(1,0,0)$ 
& 9$(1,0,0)$ 
& 10$(0,1,0)$ 
& 11$(0,\bar{1},0)$ 
& 12$(\bar{1},0,0)$ \\
$i k_1 \tau_{22} + i k_2 \tau_{21}$   
& 5$(\bar{1},0,0)$ 
& 6$(0,\bar{1},0)$ 
& 7$(0,1,0)$ 
& 8$(1,0,0)$ 
& 13$(1,0,0)$ 
& 14$(0,1,0)$ 
& 15$(0,\bar{1},0)$ 
& 16$(\bar{1},0,0)$ \\
$k_1 \tau_{21}^{'} - k_2 \tau_{22}^{'}$   
& 1$(0,1,0)$ 
& 4$(0,1,0)$ 
& 6$(1,0,0)$ 
& 7$(1,0,0)$ 
& 9$(0,\bar{1},0)$ 
& 12$(0,\bar{1},0)$ 
& 14$(\bar{1},0,0)$ 
& 15$(\bar{1},0,0)$ \\
$i k_1 \tau_{21}^{'} + i k_2 \tau_{22}^{'}$   
& 2$(1,0,0)$ 
& 3$(1,0,0)$ 
& 5$(0,\bar{1},0)$ 
& 8$(0,\bar{1},0)$ 
& 10$(\bar{1},0,0)$ 
& 11$(\bar{1},0,0)$ 
& 13$(0,1,0)$ 
& 16$(0,1,0)$ \\
$k_1 \tau_{22}^{'} - k_2 \tau_{21}^{'}$   
& 1$(0,1,0)$ 
& 2$(\bar{1},0,0)$ 
& 3$(1,0,0)$ 
& 4$(0,\bar{1},0)$ 
& 9$(0,\bar{1},0)$ 
& 10$(1,0,0)$ 
& 11$(\bar{1},0,0)$ 
& 12$(0,1,0)$ \\
$i k_1 \tau_{22}^{'} + i k_2 \tau_{21}^{'}$   
& 5$(0,\bar{1},0)$ 
& 6$(1,0,0)$ 
& 7$(\bar{1},0,0)$ 
& 8$(0,1,0)$ 
& 13$(0,1,0)$ 
& 14$(\bar{1},0,0)$ 
& 15$(1,0,0)$ 
& 16$(0,\bar{1},0)$ \\
$k_1 \tau_{21}^{''} - k_2 \tau_{22}^{''}$   
& 1$(0,0,1)$ 
& 4$(0,0,\bar{1})$ 
& 6$(0,0,\bar{1})$ 
& 7$(0,0,1)$ 
& 9$(0,0,\bar{1})$ 
& 12$(0,0,1)$ 
& 14$(0,0,1)$ 
& 15$(0,0,\bar{1})$  \\
$i k_1 \tau_{21}^{''} + i k_2 \tau_{22}^{''}$   
& 2$(0,0,\bar{1})$ 
& 3$(0,0,1)$ 
& 5$(0,0,\bar{1})$ 
& 8$(0,0,1)$ 
& 10$(0,0,1)$ 
& 11$(0,0,\bar{1})$ 
& 13$(0,0,1)$ 
& 16$(0,0,\bar{1})$  \\
$k_1 \tau_{22}^{''} - k_2 \tau_{21}^{''}$   
& 1$(0,0,1)$ 
& 2$(0,0,1)$ 
& 3$(0,0,1)$ 
& 4$(0,0,1)$ 
& 9$(0,0,\bar{1})$ 
& 10$(0,0,\bar{1})$ 
& 11$(0,0,\bar{1})$ 
& 12$(0,0,\bar{1})$ \\
$i k_1 \tau_{22}^{''} + i k_2 \tau_{21}^{''}$   
& 5$(0,0,\bar{1})$ 
& 6$(0,0,\bar{1})$ 
& 7$(0,0,\bar{1})$ 
& 8$(0,0,\bar{1})$ 
& 13$(0,0,1)$ 
& 14$(0,0,1)$ 
& 15$(0,0,1)$ 
& 16$(0,0,1)$ 
\end{tabular}
\end{table}

\end{widetext}

\begin{table}[htb]
\caption{The basis functions of the irreducible representation
of group $Fd\bar{3}m (O_h^7)$ for the two-arm star of the wave vector
${\bf k}=(\frac{1}{2},\frac{1}{2},\frac{1}{2})$. 
}
\begin{tabular}{cccccc}
Rep.
& $\psi^{{\bf k}\tau}_1$
& $\psi^{{\bf k}\tau}_2$
& $\psi^{{\bf k}\tau}_3$
& $\psi^{{\bf k}\tau}_4$
\\\hline
$\tau_{1}$ & $(0,0,0)$ & $(0,1,\bar{1})$ & $(\bar{1},0,1)$ & $(1,\bar{1},0)$ \\
\hline
$\tau_{2}$ & $(1,1,1)$ & $(0,0,0)$ & $(0,0,0)$ & $(0,0,0)$ \\
\hline
$\tau_{3}$ & $(0,0,0)$ & $(1,0,0)$ & $(0,1,0)$ & $(0,0,1)$ \\
 & $(0,0,0)$ & $(0,1,1)$ & $(1,0,1)$ & $(1,1,0)$ \\
\hline
$\tau_{5}$ & $(0,0,0)$ & $(2,0,0)$ & $(0,\bar{1},0)$ & $(0,0,\bar{1})$ \\
& $(0,0,0)$ & $(0,2,0)$ & $(0,0,\bar{1})$ & $(\bar{1},0,0)$ \\
& $(0,0,0)$ & $(0,0,2)$ & $(\bar{1},0,0)$ & $(0,\bar{1},0)$ \\
\hline
& $(0,0,0)$ & $(\bar{1},0,0)$ & $(0,2,0)$ & $(0,0,\bar{1})$ \\
& $(0,0,0)$ & $(0,0,\bar{1})$ & $(2,0,0)$ & $(0,\bar{1},0)$ \\
& $(0,0,0)$ & $(0,\bar{1},0)$ & $(0,0,2)$ & $(\bar{1},0,0)$ \\
\hline
& $(0,0,0)$ & $(1,0,0)$ & $(0,1,0)$ & $(0,0,\bar{2})$ \\
& $(0,0,0)$ & $(0,1,0)$ & $(0,0,1)$ & $(\bar{2},0,0)$ \\
& $(0,0,0)$ & $(0,0,1)$ & $(1,0,0)$ & $(0,\bar{2},0)$ \\
\hline
$\tau_{6}$ & $(2,\bar{1},\bar{1})$ & $(0,0,0)$ & $(0,0,0)$ & $(0,0,0)$ \\
 & $(0,1,\bar{1})$ & $(0,0,0)$ & $(0,0,0)$ & $(0,0,0)$ \\
\end{tabular}
\end{table}

\begin{table}[htb]
\caption{Goodness of the fit of the different spin models with ${\bf  k}= (\frac{1}{2},\frac{1}{2},\frac{1}{2})$ to the $\{\frac{1}{2},\frac{1}{2},\frac{1}{2}\}$ reflections obtained from different $\rm
ZnCr_{2}O_4$ polycrystalline samples. The best fit was obtained with the $\tau_1 + \tau_2$ model (see Fig. 10 (a)).}

\begin{tabular}{ccccc}
Sample & Rep. & T (K) & $\chi^2$ & $R_{wp}$ 
\\\hline
1   &  $\tau_1$                  & 1.5   &4.915 & .0664 \\
    & $\tau_1$,$\tau_2$         &       &4.888 & .0662 \\
    & $\tau_1$,$\tau_2$,$\tau_3$&       &4.911 & .0664 \\
    & $\tau_1$,$\tau_6$         &       &4.998 & .067  \\
    & $\tau_2$,$\tau_3$         &       &4.995 & .067  \\
2   & $\tau_1$,$\tau_2$         & 7.3   &1.368 & .0618 \\
3   & $\tau_1$,$\tau_2$         & 2     &4.003 & .0592 \\

\end{tabular}
\end{table}


\begin{references} 
\bibitem{zcoprl} 
S.-H. Lee, C. Broholm, T.H. Kim, W. Ratcliff II, and S-W. Cheong,
Phys. Rev. Lett. {\bf 84}, 3718 (2000). 
 
\bibitem{anderson}  
P.W. Anderson {\it et al}, Philos. Mag. {\bf 25}, 1 (1972). 
 
\bibitem{villain}  
J. Villain, Z. Phys. B {\bf 33}, 31 (1979). 
 
\bibitem{ramirez}  
A.P. Ramirez, ``Geometrical Frustration'' to appear in handbook 
on magnetism (2000). 
 
 
\bibitem{chubukov} 
A. Chubukov, Phys. Rev. Lett. {\bf 69}, 832 (1992). 

\bibitem{huse} 
D. A. Huse and A. D. Rutenberg Phys. Rev. B {\bf 45}, 7536 (1992). 
 
\bibitem{mila98}  
F. Mila, Phys. Rev. Lett. {\bf 81}, 2356 (1998). 
 
\bibitem{obradors}  
X. Obradors, A. Labarta, A. Isalgue, J. Tejada, J. Rodriguez, 
and M. Pernet, Solid State Commun. {\bf 65} 189 (1988).

\bibitem{aprscgo90}  
A.P. Ramirez, G.P. Espinosa, and A. S. Cooper, 
Phys. Rev. Lett. {\bf 64}, 2070 (1990). 

\bibitem{cb90}  
C. Broholm, G. Aeppli, G.P. Espinosa, and A.S. Cooper,
Phys. Rev. Lett. {\bf 65}, 3173 (1990). 
 
\bibitem{scgoeuro} 
S.-H. Lee, C. Broholm, G. Aeppli, A.P. Ramirez, T.G. Perring,
C.J. Carlile, M. Adams, T.J.L. Jones, and B. Hessen, Europhys. Lett. {\bf 35}(2), 127 (1996) 

\bibitem{park} K. Park  and S. Sachdev, Phys. Rev. B {\bf 65}, 220405(R) (2002).
\bibitem{senthil} T. Senthil, L. Balents, S. Sachdev, A. Vishwanath, M. P. A. Fisher,  Phys. Rev. B {\bf 70}, 144407 (2004).
\bibitem{ran} Y. Ran, M. Hermele, P. A. Lee, and X.-G. Wen, cond-mat/0611414.
\bibitem{ryu} S. Ryu, O. I. Motrunich, J. Alicea, and M. P. A. Fisher, cond-mat/0701020.
\bibitem{matan} K. Matan, D. Grohol, D.G. Nocera, Y. Yildirim, A.B. Harris, S.-H. Lee, S.E. Nagler, Y.S. Lee, Phys. Rev. Lett. {\bf 96}, 247201 (2006).
\bibitem{shores} M. P. Shores, et al., J. Am. Chem. Soc. {\bf 127}, 13462-13463 (2005).
\bibitem{helton} J. S. Helton et al., Phys. Rev. Lett. {\bf 98}, 107204 (2007).
\bibitem{ofer} O. Ofer et al., Phys. Rev. Lett. {\bf ??}, ??.
\bibitem{mendels} P. Mendels et al., Phys. Rev. Lett. {\bf 98}, 077204 (2007).
\bibitem{shlnmat} S.-H. Lee, H. Kikuchi, Y. Qiu, B. Lake, Q. Huang,  K. Habicht, and K. Kiefer, Nature Materials, in press (2007).

\bibitem{cana98}  
B. Canals and C. Lacroix, Phys. Rev. Lett. {\bf 80}, 2933 (1998);
Phys. Rev. B {\bf 61}, 1149 (2000).
 
 \bibitem{moes98}  
R. Moessner and J.T. Chalker, Phys. Rev. Lett. {\bf 80}, 2929 (1998) ;
Phys. Rev. B {\bf 58}, 12049 (1998). 

\bibitem{zconature} 
S.-H. Lee, C. Broholm, W. Ratcliff, G. Gasparovic, Q. Huang, T. H. Kim, and S-W. Cheong,
Nature {\bf 418}, 856 (2002). 

\bibitem{cdcr2o4} J.-H. Chung, M. Matsuda, S.-H. Lee, K. Kakurai, H. Ueda, T.J. Sato, H. Takagi, K.-P. Hong, and S. Park, Phys. Rev. Lett. {\bf 95}, 247204 (2005). 

\bibitem{hgcr2o4} M. Matsuda, H. Ueda, A. Kikkawa, Y. Tanaka, K. Katsumata, Y. Narumi, T. Inami, Y. Ueda, S.-H. Lee, Nature Physics {\bf 3}, 397 (2007).

\bibitem{good}  
J.B. Goodenough, Phys. Rev. {\bf 117}, 1442 (1960).  

\bibitem{tsunetsugu} H. Tsunetsugu and Y. Motome, Phys. Rev. B {\bf 68}, 060405(R) (2003).

\bibitem{znv2o4} S.-H. Lee, D. Louca, T. Sato, Y. Ueda, M. Isobe, S. Rosenkranz, and R. Osborn, Phys. Rev. Lett. {\bf 93}, 156407 (2004). 

\bibitem{oleg_znv2o4} O. Tchernyshyov, Phys. Rev. Lett. {\bf 93}, 157206 (2004).

\bibitem{khomskii} D. I. Khomskii and T. Mizokawa, Phys. Rev. Lett. {\bf 94}, 156402 (2005).

\bibitem{foroleg}  The formula of the hexagonal magnetic structure factor was first derived by O. Tchernyshyov {\it et al.} in the different context of spin waves in long range ordered phases on the spinel lattice. There it accounts for the structure factor of a nonzero-energy local antiferromagnon on {\it ferromagnetically} aligned hexagonal spin loops. See O. Tchernyshyov, R. Moessner, and S. L. Sondhi, Phys. Rev. Lett. {\bf 88}, 067203 (2002). 

\bibitem{acr2o4_cryst} S.-H. Lee, G. Gasparovic, C. Broholm, M. Matsuda, J.-H. Chung, Y. J. Kim, H. Ueda, G. Xu, P. Zschack, K. Kakurai, H. Takagi, W. Ratcliff, T. H. Kim, and S-W. Cheong, J. of Phys. Ð Cond. Matt. {\bf 19 (14)}, 145259 (2007).

\bibitem{hueda} H. Ueda, H. Mitamura, T. Goto, and Y. Ueda, Phys. Rev. B {\bf 73}, 094415 (2006).

\bibitem{hueda_prl} H. Ueda, H. Aruga-Katori, H. Mitamura, T. Goto, and H. Takagi, Phys. Rev. Lett. {\bf 94}, 047202 (2005).

\bibitem{penc} K. Penc, N. Shannon, and H. Shiba, Phys. Rev. Lett. {\bf 93}, 197203 (2004).

\bibitem{balents} D. L. Bergman, R. Shindou, G. A. Fiete, and L. Balents, Phys. Rev. Lett. {\bf 96}, 097207 (2006); Phys. Rev. B {\bf ??} cond-mat/0605467 (2006).

\bibitem{ks} Due to the very weak tetragonal distortion, it is not easy to determine if the half integer wavevector of ${\bf k} = (1,0,\frac{1}{2})$ and the plane of the ${\bf k} = (1/2,1/2,0)$ is parallel with or perpendicular to the contracted $c$-axis. Recently, however, neutron diffraction measurements with uniaxial pressure along a few different directions proved \cite{zcojpjs} that the half integer of ${\bf k} = (1,0,\frac{1}{2})$ is along the $c$-axis and the ${\bf k}=(1/2,1/2,0)$ is perpendicular to the $c$-axis.

\bibitem{zcojpjs} 
I. Kagomiya, Y. Hata, D. Eto, H. Yanagihara, E. Kita, K. Nakajima, K. Kakurai, M. Nishi, and K. Ohoyama, J. Phys. Soc. Jpn {\bf 76}, 064710 (2007). 

\bibitem{fiorani84}  
D. Fiorani, S. Viticoli, J. L. Dormann, J. L. Tholence,
and A. P. Murani, Phys. Rev. B {\bf 30}, 2776 (1984). 

\bibitem{percol}  
F. Scholl and K. Binder, Z. Phys. B {\bf 39}, 239 (1980).

\bibitem{hammann}  
J. Hammann, D. Fiorani, M. El Yamani, and J. L.Dormann, 
J. Phys. C: Solid State Phys. {\bf 19}, 6635-6644 (1986). 

\bibitem{kcodsoprb} S.-H. Lee, C. Broholm, M.F. Collins, L. Heller,
A.P. Ramirez, Ch. Kloc, E. Bucher, R.W. Erwin, N. Lacevic, 
Phys. Rev. B {\bf 56}, 8091 (1997).

\bibitem{polcor}
C.F. Majkrzak, Physica B {\bf 221}, 342 (1996).

\bibitem{oles70}  
A. Ol\'{e}s, Phys. Status Solidi A {\bf 3}, 569 (1970).
\bibitem{shaked70}  
H. Shaked, J. M. Hastings, and L. M. Corliss, Phys. Rev. B {\bf 1}, 3116 (1970).

\bibitem{moon} R.M. Moon {\it et al.}, Phys. Rev. {\bf 181},
920 (1969). 

\bibitem{lovesey}  
S.M. Lovesey, {\it Theory of Thermal Neutron Scattering from 
Condensed Matter}, (Clarendon Press, Oxford) 1984. 


\bibitem{schiessl} W. Schiessl, W. Potzel, H. Karzel, M. Steiner,
G.M. Kalvius, A. Martin, M.K. Krause, I. Halevy, J. Gal, W. sch\"{a}fer,
G. Will, M. Hillberg, and R. W\"{a}ppling, Phys. Rev. B {\bf 53}, 9143 (1996).

  

\bibitem{izyumov}
Yu. A. Izyumov, V. E. Naish, and R. P. Ozerov, {\it Neutron Diffraction
of Magnetic Materials}, (Plenum Publishing Corporation, New York) 1991.

 


 
\bibitem{xraytables}P. J. Brown, in ``International Tables for  
Crystallography'', Volume C, edited by A. J. C. Wilson and E.Prince, Kluwer Academic Publishers Boston (1999). 

\bibitem{kova}
O.V. Kovalev, {\it Representations of the Crystallographic Space Groups},
(Gordon and Breach Science Publishers) 1993.

\bibitem{furr79}
A. Furrer and H. U. G\"{u}del, J. Magn. Magn. Mater. {\bf 14}, 256 (1979).

\bibitem{moeber99} R. Moessner and A. J. Berlinsky, Phys. Rev. Lett. {\bf 83}, 3293 (1999).

\bibitem{scgoprl} 
S.-H. Lee, C. Broholm, G. Aeppli, T.G. Perring, B. Hessen, and A. Taylor,
Phys. Rev. Lett. {\bf 76}, 4424 (1996). 

\bibitem{apr92} A.P. Ramirez, G.P. Espinosa, and A.S. Cooper, 
Phys. Rev. B {\bf 45}, 2505 (1992).

\bibitem{schiffer} P. Schiffer and I. Daruka, Phys. Rev. B {\bf 56}, 13712 (1997).

 
 

 
\end{references}
\end{document}